\documentclass{article}
\usepackage{arxiv}
\usepackage[utf8]{inputenc} 
\usepackage[T1]{fontenc}    
\usepackage{hyperref}       
\usepackage{url}            
\usepackage{booktabs}       
\usepackage{amsfonts}       
\usepackage{nicefrac}       
\usepackage{microtype}      
\usepackage{lipsum}		
\usepackage{graphicx}
\usepackage[numbers]{natbib}
\usepackage{doi}
\usepackage{array,ragged2e,longtable}
\usepackage{amsmath}
\usepackage{ulem}
\usepackage{comment}
\usepackage{multirow}
\usepackage{booktabs}

\title{Temporal Patterns of Multiple Long-Term Conditions in Individuals with Intellectual Disability Living in Wales: An Unsupervised Clustering Approach to Disease Trajectories}

\author{ \href{https://orcid.org/0000-0002-9883-0381}{\includegraphics[scale=0.06]{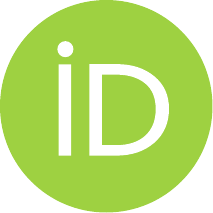}\hspace{1mm}Rania~Kousovista}\\
	Department of Computer Science\\
	Loughborough University\\
	Loughborough, United Kingdom, LE11 3TU\\
	\And
	\href{https://orcid.org/0000-0002-4663-6907}{\includegraphics[scale=0.06]{orcid.pdf}\hspace{1mm}Georgina~Cosma}\thanks{Corresponding Author: g.cosma@lboro.ac.uk} \\
	Department of Computer Science\\
	Loughborough University\\
	Loughborough, United Kingdom, LE11 3TU\\
	\And
    \href{https://orcid.org/0000-0002-4742-3102}{\includegraphics[scale=0.06]{orcid.pdf}\hspace{1mm}Emeka~Abakasanga}
    \\
	Department of Computer Science\\
	Loughborough University\\
	Loughborough, United Kingdom, LE11 3TU\\	
	\And
	\href{https://orcid.org/0000-0003-0814-0801}{\includegraphics[scale=0.06]{orcid.pdf}\hspace{1mm}Ashley Akbari} \\
    Faculty of Medicine\\
    Health and Life Science\\ Swansea University, Swansea, United Kingdom \\
	\And
	Francesco Zaccardi  \\
	Leicester Diabetes Center, University of Leicester\\
    Leicester, United Kingdom
    \And
	Gyuchan Thomas Jun \\
	School of Design and Creative Arts\\
	Loughborough University\\
	Loughborough, United Kingdom
	\And
	Reza Kiani \\Leicestershire Partnership NHS Trust \\ and Leicester University\\ Leicester, United Kingdom
    \And
	Satheesh Gangadharan\\Leicestershire Partnership NHS Trust\\ Leicester, United Kingdom
}

\date{}

\renewcommand{\shorttitle}{Temporal Patterns of Long-Term Conditions in Intellectual Disabilities}

\hypersetup{
pdftitle={Temporal Patterns of Long-Term Conditions in Intellectual Disabilities},
pdfsubject={cs.AI},
pdfauthor={Rania~Kousovista, Georgina~Cosma, Emeka~ Abakasanga Ashley~Akbari, Francesco~Zacardi, Gyuchan Thomas~Jun, Reza~Kiani, Satheesh~Gangadharan},
pdfkeywords={Disease trajectories, Chronic disease, Co-morbidity, Clustering, Intellectual Disabilities},
}

\begin{document}
\maketitle

\begin{abstract}
Identifying and understanding the co-occurrence of multiple long-term conditions (MLTC) in individuals with intellectual disabilities (ID) is crucial for effective healthcare management. These individuals often experience earlier onset and higher prevalence of MLTCs compared to the general population, highlighting the urgency of this issue. Despite established research on the high prevalence of MLTCs in the ID population,  the specific patterns of co-occurrence and temporal progression of these conditions remain largely unexplored. This study presents an innovative unsupervised approach for examining and characterising clusters of MLTC in individuals with ID, based on their shared disease trajectories. Using a dataset of electronic health records (EHRs) from 13069 individuals with intellectual disabilities, encompassing primary and secondary care data in Wales from 2000 to 2021, this study analyses the time sequences of ordered disease diagnoses. The study population comprised 52.3\% males and 47.7\% females, with a mean of 4.5 $\pm$ 3 long-term conditions (LTCs) per patient. Distinct MLTC clusters were identified in both males and females, stratified by age groups ($<$ 45 and $\geq$ 45 years). For males under 45, a single cluster dominated by neurological conditions (32.4\%), while three clusters were identified for older males, with the largest characterised by circulatory (51.8\%). In females under 45, one cluster was found with digestive system conditions (24.6\%) being most prevalent. For females $\geq$ 45 years, two clusters were identified: the first cluster was predominantly defined by circulatory (34.1\%), while the second cluster by digestive (25.9\%) and musculoskeletal (21.9\%) system conditions. Mental illness, epilepsy, and reflux disorders were prevalent across all groups. This study reveals complex multimorbidity patterns in individuals with ID, highlighting age and sex differences. The identified clusters provide new insights into disease progression and co-occurrence in this population. These findings can inform the development of targeted interventions and risk stratification strategies, potentially improving personalised healthcare for individuals with ID and MLTCs with the aim of improving health outcome for this vulnerable group of patients, i.e., reducing frequency and length of hospital admissions and premature mortality.
\end{abstract}


\section{Introduction}
People with intellectual disability (ID) face a significantly higher risk of developing a range of physical and mental health conditions compared to the general population. These conditions often occur at a younger age and lead to poorer outcomes, owing to a combination of genetic, behavioural, and social factors \citep{cooper2020rates, kinnear2018prevalence, emerson2014health}. Studies show a much higher occurrence of multiple long-term conditions (MLTCs) in this population \citep{kinnear2018prevalence}. MLTCs, defined as two or more conditions in addition to ID, is linked to premature death and poorer quality of life \citep{heslop2018learning}. Despite this, there appear to be only a few studies reporting the prevalence of MLTCs conducted on a large scale \citep{kinnear2018prevalence,carey2016health}, but no studies were found to reveal patterns of MLTCs and conditions more likely to co-occur together in this population.\\
The growing use of electronic health records (EHRs) has enabled significant advances in addressing clinical challenges, enhancing diagnostic capabilities, and improving patient outcomes \cite{blumenthal2010meaningful, solares2020deep, pham2016deepcare}. In addition to enabling studies on co-occurring conditions, the longitudinal nature of EHR provides a unique opportunity to uncover temporal associations and trajectories between conditions. Importantly, chronic health conditions frequently co-occur more than expected by chance, often as a consequence of shared risk factors, pathogenicity, or their treatment \cite{johnston2019defining}. 
However, most prior studies have not incorporated the time dimension due to the short time span of the available data \citep{chmiel2014spreading, teno2001dying}. \\
Only recently have a few large-scale analyses assessed disease trajectories by evaluating temporal ordering of co-morbidity pairs over time in general population \citep{jensen2014temporal, giannoula2018identifying, lyons2023trajectories}. Many studies further developed the framework initially proposed by Jensen et al. \citep{jensen2014temporal}, who described general principles for temporal trajectory analysis using Danish national data. For instance, Siggaard et al. \citep{siggaard2020disease} published a browser of these results, while Jørgensen and Brunak \citep{jorgensen2021time} focused on chronic obstructive pulmonary disease (COPD) trajectories. Hu et al. \citep{hu2019large} linked the data to a cancer registry to investigate pre-diagnosis trajectories. Jensen et al.'s \cite{jensen2014temporal} approach has been applied, with modifications, to other populations including post-depression trajectories in UK Biobank \citep{han2021disease}, and end-of-life trajectories in California \citep{paik2021condensed}. Furthermore, Giannoula et al. \citep{giannoula2018identifying,giannoula2021system} proposed a framework to detect and cluster co-morbidity pairs and shared trajectories over time using a dynamic time warping (DTW)-based unsupervised algorithm in EHRs, and later extended this to include genetic information in the clustering step. Unsupervised algorithms discover natural patterns in data without learning predefined outcomes or classifications. Trajectory analyses can reveal complex, time-ordered condition associations, as well as MLTCs patterns to enable better understanding of disease progression for improved prediction outcomes.\\
In this study, we propose a computational framework for the analysis of temporal MLTCs on EHRs in 13069 adults diagnosed with ID in Wales, incorporating 40 long-term conditions (LTCs) from both primary and secondary care data. While most prior studies have applied temporal trajectory analysis to secondary care data and ICD-9 or -10 codes \citep{jensen2014temporal, giannoula2018identifying, siggaard2020disease, jorgensen2021time, hu2019large, han2021disease, paik2021condensed, giannoula2021system}, with only one study using primary care data \citep{planell2020trajectories}, our approach utilises both to fully capture MLTCs, as most chronic conditions are treated in general practice. This approach highlights several differences in MLTC patterns between male and female sub-populations across different age groups, acknowledging sex and age as crucial factors in understanding MLTCs. The primary contributions of this research include statistical analysis to identify significant temporal condition pairs, identification of shared MLTCs trajectories, construction of a network of all shared trajectories, and identification of trajectory clusters using an unsupervised machine learning algorithm.

\section{Materials and Methods}
\subsection*{Study Design and Participants} 
\noindent This study focused on Welsh residents aged 18 and older with intellectual disability and at least one long-term condition (LTC) between 1st January 2000, and 31st December 2021. This population-based study utilised the Secure Anonymised Information Linkage (SAIL) Databank, a Welsh data repository that enables individual-based data linkage across datasets \citep{lyons2009sail}. We identified eligible individuals who were registered with a general practitioner (GP) at the study start date (Figure \ref{fig:flowchart}). For inclusion in the cohort, individuals required key identifying information as defined within Wales, including a unique anonymised patient identifier, age (or date of birth), sex, residential (WIMD) and GP registration information. Primary and secondary care electronic health records (EHRs) securely stored within SAIL were used to capture LTC diagnoses. Demographic data were used from the Welsh Demographic Service Dataset (WDSD) that contains information relating to people who are resident in Wales and registered with a Welsh GP. Data collected by GPs is captured via Read v2 codes (5-digit codes related to diagnosis, medication, and process of care codes). Hospital in- and out- patient data are collected in the Patient Episode Database for Wales, which contains clinical information regarding patients’ hospital admissions, discharges, diagnoses and operations utilising the International Classification of Diseases (ICD-10) clinical coding system. The Annual District Death Extract (ADDE) from the Office for National Statistics (ONS) was used to capture all deaths and dates of death that occurred over the study period for all Welsh residents, contains information regarding the dates and causes of deaths (also ICD-10).\\
In this study, a LTC is defined as a condition that cannot, at present, be cured but is controlled by medication and/or other treatment/therapies \citep{DepartmentOfHealthAndSocialCare2012}. For conditions that do not always fall into the chronic category, we applied duration-based criteria to define them as long-term or chronic (Figure \ref{fig:duration}). For the purpose of this study, multiple long-term conditions (MLTCs) were defined as two or more chronic conditions. We selected 40 LTCs for this study (Table \ref{tab:ltc_list}), based on consensus from a multidisciplinary professional advisory panel. The professional advisory panel (PAP) comprising a team of experts, including General Practitioners, a consultant Psychiatrist, nurses, pharmacists, and data analysts. The full details of condition merging, grouping, and the comprehensive list of Read v2 and ICD-10 codes for each condition can be found in our study protocol \cite{Shabnam_Kousovista_Abakasanga_Kaur_Cosma_Akbari_Gangadharan_Zaccardi_2024}. 

\subsection*{Data Analysis and Computational Methods}
\noindent We introduced a novel methodology for identifying and analysing shared disease trajectories in patients with ID. Our approach comprises three main stages: (1) identify pairwise condition associations and their temporal directions; (2) construct shared MLTC trajectories; and (3) utilise a network-based technique to cluster these trajectories into meaningful clusters of similar disease trajectories. 

\paragraph*{Identifying the Pairwise Condition Associations and Their Temporal Direction}
\noindent The extracted primary (i.e., GP) and secondary (i.e., hospital) data for all patients were harmonised to a single table, where each row represents a unique patient and each unique variable column represents a binary indicator of the patient's diagnosis of one of the LTCs. For every patient, we extracted the date of first diagnosis for each LTC, creating a chronological sequence of LTC diagnoses. The analysis was stratified according to sex and age. Age was categorised into two groups, under 45 years old ($<$45) and 45 years old and above ($\geq 45$), taking into account the median observed age per patient's clinical history. \\
Thereafter, we derived all possible pairwise combinations of LTCs from our dataset, considering only those where at least 10 patients shared both conditions and had a minimum temporal separation of six months between the diagnoses. Fisher's exact test was then implemented on 2 × 2 contingency tables constructed for each qualifying pair of conditions. The resulting p-values were then corrected using the Bonferroni correction for multiple testing, with a threshold of $\alpha$ = 0.001. For all co-morbidity pairs that demonstrated a significant association, we assessed whether a statistically significant temporal order (direction) existed between condition 1 ($C_1$) and condition 2 ($C_2$). Specifically, a Binomial test was used to evaluate the temporal direction of diagnoses, comparing the number of patients for whom condition $C_2$ follows condition $C_1$ against those where $C_1$ follows $C_2$ ($C_1$ → $C_2$ versus $C_2$ → $C_1$). If the p-value $< 0.05$ from the Binomial test, a preferred direction was assigned to that pair of conditions, based on the more frequently occurring sequence.

\paragraph*{Identifying the Shared MLTC Trajectories}
\noindent Shared disease trajectories were developed by combining significant temporally ordered condition pairs into longer sequences of MLTCs. These pairs were combined to form all possible longer trajectories. For example, if the pairs $C_1$ → $C_2$ and $C_2$ → $C_3$ were found to be significant, they were combined to form the trajectory $C_1$ → $C_2$ → $C_3$. Pairs with statistically significant (preferred) directionality were included (p-values $<$ 0.05 in the Binomial test for directionality), while in the case of no preferred directionality (p-values $\geq$ 0.05), both directions were considered. The actual occurrences of these trajectories in the patient population were then counted. These trajectories could contain other intermediate conditions, as long as the conditions maintained their significant chronological order. Consequently, a list of all identified trajectories was obtained, along with their respective occurrence counts in the patient cohort. These trajectories varied in length, with the longer ones representing sequences of conditions where two or more patients shared the exact same chronological order. All trajectories with a length of three conditions shared in more than ten patients used for clustering. For trajectories containing the same three conditions, only the most frequent unique sequence was retained, ensuring each set of three conditions was represented by its most common temporal order.

\paragraph*{Identifying Trajectory Clusters}
\noindent We propose a network-based clustering technique that employs a novel approach to quantify and analyse the associations between conditions across multiple disease trajectories. This method consists of four key steps: constructing a trajectory condition network, developing a condition similarity metric based on shortest paths, creating a trajectory similarity matrix, and applying spectral clustering. These steps are described as follows.\\
\begin{figure}[t]
    \centering
     \includegraphics[width=\textwidth]{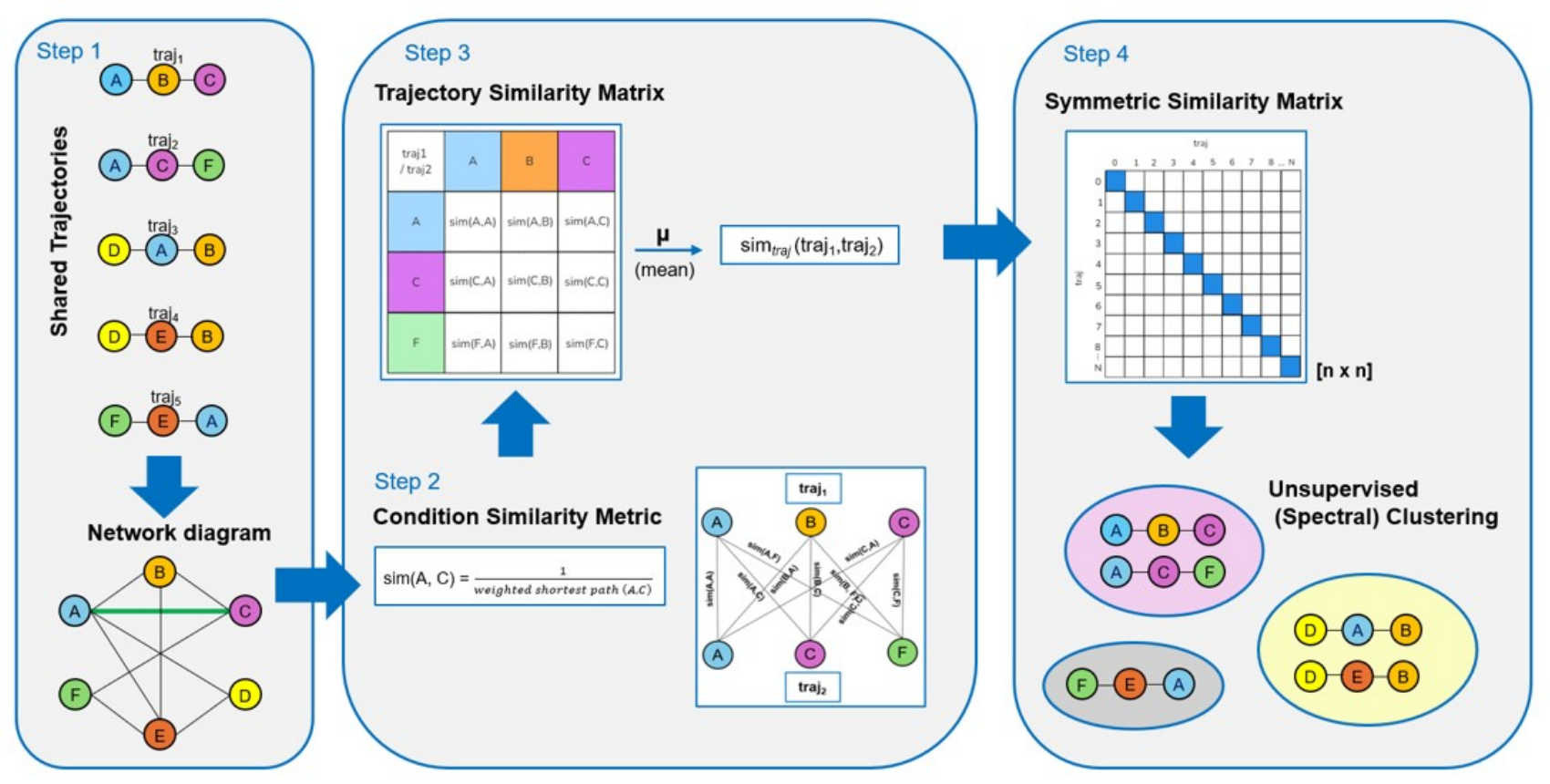}
         \caption{Schematic representation of the workflow of the proposed methodology.}
         \label{fig:methodology}
\end{figure}
\textbf{Step 1.} The first step is about constructing a \textit{trajectory condition network}. A network of all trajectories was constructed to explore the associations between conditions across multiple trajectories and define a similarity metric among them (Figure~\ref{fig:methodology}). Let $G = (N, E)$ be an undirected graph (or network graph) where $N$ is the set of nodes (conditions) and $E$ is the set of edges. Each edge $e_{i,j} \in E$ connects two conditions $c_i$ and $c_j$, where $i, j \in {1, 2, \ldots, k}$, where $k$ is the total number of unique conditions across all trajectories.\\
\textbf{Step 2.} To quantify similarity between trajectory conditions we developed a \textit{condition similarity metric} based on the shortest path problem in graph theory. The weight $w$ of an edge $e_{i,j}$, where $w:E\rightarrow \mathbb{R}$, is defined as:
\begin{equation} \label{weight}
w(e_{i,j}) = \frac{1}{\sqrt{f(e_{i,j})}}
\end{equation}
where $f(e_{i,j})$, $f:E\rightarrow \mathbb{R}$, is the frequency of edges between conditions $c_i$ and $c_j$ in the network. This weight function assigns lower weights to more frequent edges, effectively making frequently co-occurring conditions closer in the network graph.\\
Let $P = (c_1, c_2, ..., c_n)$ be a path from $c_1$ to $c_n$ in network graph $G$. The shortest path length is given by: 
\begin{equation} \label{shortestpathmetric}
\textit{shortest path}(c_1, c_n) = \min_{P} \sum_{l=1}^{n-1} w(e_{l,l+1})
\end{equation}
where the minimum is taken over all possible paths from $c_1$ to $c_n$ in the graph, and $w(e_{l,l+1})$ is the weight of the edge between consecutive conditions $c_l$ and $c_{l+1}$ in the path. We compute this shortest path using Dijkstra's algorithm \citep{dijkstra2022note}. To convert the shortest path lengths into similarity scores ranging from 0 to 1, we take the inverse:
\begin{equation} \label{shortestpath}
\textit{sim}(c_i, c_j) = \frac{1}{\textit{shortest path}(c_i, c_j)}
\end{equation}
This transformation ensures that conditions with shorter path lengths between them (i.e., more closely related) have higher similarity scores.\\
\textbf{Step 3.} To assess the overall similarity between trajectories, we constructed an $n \times n$ \textit{symmetric similarity matrix}, where $n$ is the total number of trajectories (Figure~\ref{fig:methodology}). 
The similarity between two trajectories, $traj_i$ and $traj_j$, is calculated as the mean of the shortest path similarities between their respective conditions. This can be expressed as:
\begin{equation} \label{eq:traj_similarity}
    sim_{traj}(traj_i, traj_j) = \frac{1}{|traj_i| \cdot |traj_j|} \sum_{c_i \in traj_i} \sum_{c_j \in traj_j} sim(c_i, c_j)
\end{equation}
where $|traj_i|$ and $|traj_j|$ are the numbers of conditions in trajectories $i$ and $j$ respectively, $i, j \in {1, 2, \ldots, n}$ where $n$ is the total number of trajectories, and $sim(c_i, c_j)$ is the shortest path similarity between conditions $c_i$ and $c_j$ as defined in Eq. \ref{shortestpath}.\\
\textbf{Step 4.} To identify clusters of highly similar trajectories based on their condition similarities, we applied spectral clustering to the precomputed trajectory similarity matrix ($sim_{traj}$). This unsupervised machine learning algorithm partitions the data by exploiting the eigenstructure of the similarity matrix. We implemented spectral clustering using the Scikit-Learn library with default parameter settings \citep{pedregosa2011scikit}. The algorithm treats the trajectory similarity matrix as a weighted graph adjacency matrix, performs a spectral embedding of the data points into a lower-dimensional space, and then clusters the embedded points using the k-means algorithm. The optimal number of clusters was determined using the Calinski-Harabasz score.

\section{Results}
\subsection*{Data Analysis}
\noindent Table \ref{tab: characteristics} summarises the counts and percentages of patient demographic characteristics included in the study and reports the mean number of LTCs, prevalence of MLTCs (\%), and prevalence of physical-mental MLTCs (\%) stratified by sex, age group, ethnic group, and Welsh Index of Multiple Deprivation (WIMD) quintiles. The study population comprised 13069 patients, with 52.3\% being male. The mean number of LTCs for all patients was 4.5 ($\pm$ 3), with 85.9\% having MLTCs and 31.8\% having physical-mental MLTCs.
Subgroup analysis revealed notable trends across demographic categories. Females showed a slightly higher mean number of LTCs (4.9 $\pm$ 3.3) and prevalence of MLTCs (87.7\%) compared to males (4.2 $\pm$ 2.8 and 84.3\% respectively) (Tables \ref{tab: characteristics} and \ref{tab:nconditions-gender}). Patients aged 45 years and older had a higher mean number of LTCs (5.2 $\pm$ 3.2) and prevalence of MLTCs (91\%) compared to those under 45 (3.5 $\pm$ 2.4 and 79\%). Socioeconomic factors, indicated patients from the most deprived areas had a higher mean number of LTCs (4.9 $\pm$ 3.2) and prevalence of MLTCs (88.9\%) compared to those from the least deprived areas (4.5 $\pm$ 2.9 and 85.6\%). 

\subsection*{Analysis of Pairwise Condition Associations}
\noindent Table \ref{tab:gndr-demo} shows the total number of patients after stratification. A comparison of the prevalence of several conditions among males and females, as well as between age groups, can be found in Supplementary Tables \ref{tab:malecondcounts} and \ref{tab:femalecondcounts}.  The statistically significant condition pairs are provided in Table \ref{tab:genderpairs}.\\

\textbf{Males}. Mental illness was the most prevalent condition affecting 32.8\% of all males (Table \ref{tab:malecondcounts}). A key finding was that its prevalence was higher in those under 45 years (35\%) compared to those 45 and above (31\%). Epilepsy followed closely, present in 31\% of all males, again with a higher prevalence in the younger group (35.3\% vs 27.7\%). Reflux disorders showed a consistent prevalence of 29.7\% across both age groups. Hypertension demonstrated a marked increase with age, affecting 15.5\% of males under 45 and 29.7\% of those 45 and above. Additionally, chronic kidney disease (CKD) showed a significant increase from 10.4\% in the younger group to 26.9\% in the older group. Similarly, diabetes increased from 13.8\% in the younger group to 25.3\% in the older group.\\ The most frequent condition pair association in males under 45 (encountered in 385 patients) was found to be between mental illness and reflux disorders (Table \ref{tab:genderpairs}). Other significant co-morbidities include pairs of different neuropsychiatric conditions (e.g., epilepsy $\rightarrow$ mental illness in 287 patients, mental illness $\rightarrow$ insomnia in 267 patients, and cerebral palsy $\leftrightarrow$ epilepsy in 205 patients), combinations of neuropsychiatric and haematological conditions (e.g., mental illness $\rightarrow$ anaemia, 267 patients) or associations between endocrine and circulatory with renal conditions (e.g., diabetes $\rightarrow$ CKD and hypertension $\leftrightarrow$ CKD in 82 patients). In males aged 45 and above, hypertension $\rightarrow$ CKD emerged as the most prevalent pair, affecting 447 patients, followed by mental illness $\leftrightarrow$ reflux disorders (422 patients) and hypertension $\rightarrow$ diabetes (411 patients). \\
\begin{table}[t] \centering
\small
\caption{Patient characteristics and description of LTC.}
\label{tab: characteristics}
\begin{tabular}{rlllp{1in}}
\toprule
\textbf{Group} & \textbf{Patients, N(\%)} & \textbf{Mean LTC ($\pm$SD)} & \textbf{Patients with MLTC (\%)}$^a$ & \textbf{Patients with physical–mental MLTC (\%)}$^a$ \\\midrule
All Patients & 13069 (100.0) & 4.5 (3.0) & 11231 (85.9) & 4162 (31.8) \\ \midrule
Sex & \multicolumn{4}{l}{}                                            \\ \midrule
Male & 6830 (52.3) & 4.2 (2.8) & 5757 (84.3) & 2055 (30.1) \\
Female & 6239 (47.7) & 4.9 (3.3) & 5474 (87.7) & 2107 (33.8) \\ 
\midrule
Age (years) &\multicolumn{4}{l}{} \\ \midrule
$<$ 45 & 5506 (42.1) & 3.5 (2.4) & 4350 (79.0)  & 1800 (32.7)  \\
$\geq$ 45 & 7563 (57.9) & 5.2 (3.2) & 6881 (91.0)  & 2362 (31.2)  \\\midrule
Ethnic group &\multicolumn{4}{l}{}    \\\midrule
White & 9161 (70.1) & 4.8 (3.1) & 8074 (88.1) & 3259 (35.6) \\
Unknown & 3638 (27.8) & 3.8 (2.7) & 2916 (80.2) & 816 (22.4) \\
Asian & 179 (1.4) & 4.6 (2.7) & 167 (93.3) & 60 (33.5) \\
Black & 43 (0.3) & 3.8 (2.9) & 35 (81.4) & 14 (32.6) \\
Mixed & 15 (0.1) & 3.7 (2.7) & 12 (80.0) & 5 (33.3) \\
Other & 33 (0.3) & 4.1 (3.1) & 27 (81.8) & 8 (24.2)\\\midrule
WIMD & \multicolumn{4}{l}{} \\ \midrule
1. Most deprived & 3195 (24.4) & 4.9 (3.2) & 2840 (88.9) & 1148 (35.9) \\ 
2 & 2682 (19.8) & 4.6 (3.1) & 2232 (86.4)  & 834 (32.3) \\ 
3 & 2567 (19.6) & 4.4 (3.0) & 1797 (86.1) & 597 (28.6) \\ 
4 & 1949 (14.9) & 4.5 (2.9) & 1699 (87.2) & 564 (28.9) \\
5. Least deprived & 1346 (10.3) & 4.5 (2.9) & 1152 (85.6) & 379 (28.2) \\
\bottomrule
\end{tabular}
\par\smallskip
\tiny
\noindent $^a$Proportions calculated using the corresponding subgroup number in ``Patients, N(\%)'' column as the denominator. LTC = Long-term condition; SD = standard deviation; MLTC = Multiple long-term condition; Physical-mental MLTC = having at least one physical and one mental health condition. WIMD = Welsh Index of Multiple Deprivation quintiles categories range from 1 (most deprived) to 5 (least deprived).
\end{table}\\
\textbf{Females}. Mental illness was also the most common condition among females, affecting 35.1\% overall (Table \ref{tab:femalecondcounts}). 
Thyroid conditions were prevalent in 30.8\% of all females, with 33.3\% in the 45 and above age group and 27.1\% in those under 45. Reflux disorders affected 30.1\% of all females, with 30.9\% in the under 45 group and 29.5\% in those 45 and above. Hypertension showed a significant age-related increase, from 11.9\% in those under 45 to 31.9\% in those 45 and above. As expected, menopausal and perimenopausal conditions were more prevalent in the older female group, affecting 25.9\% of those 45 and above compared to 8.8\% in those under 45.\\
The most frequent condition pair in females under 45 was mental illness $\rightarrow$ reflux disorders, shared by 375 patients (Table \ref{tab:genderpairs}). Chronic airway conditions featured prominently in this group, with chronic airway diseases $\rightarrow$ mental illness and chronic airway diseases $\rightarrow$ reflux disorders affecting 357 and 293 patients, respectively. In the older age groups, we noted significant associations between circulatory and musculoskeletal conditions (e.g., hypertension $\rightarrow$ chronic arthritis, 484 patients), followed by mental illness $\rightarrow$ reflux disorders (473 patients) and mental illness $\rightarrow$ chronic arthritis (470 patients).\\
In general, the younger age groups, particularly those under 45, showed a higher prevalence of co-morbidities involving neuropsychiatric conditions, often in combination with other system conditions (Table \ref{tab:genderpairs}). This was evident in both males and females, with mental illness frequently co-occurring with conditions from various other physiological systems. While mental illness and epilepsy were among the top five most prevalent conditions in both sexes, thyroid disorders ranked second in females but did not appear in the top five for males (Tables \ref{tab:malecondcounts} and \ref{tab:femalecondcounts}). Hypertension was among the top five conditions in males across all ages, but in females, it only ranked in the top five for the 45 and above age group.

\begin{table}[!t]
\caption{Top co-morbidity pairs in males and females by age group.}
\label{tab:genderpairs}
\tiny
\begin{tabular}{lccclcc}
\toprule
\multicolumn{3}{c}{\textbf{Males $<$ 45 years}} && \multicolumn{3}{c}{\textbf{Males $\geq$ 45 years}} \\
\cmidrule{1-3} \cmidrule{5-7}
\textbf{Condition Pairs} & \textbf{N Patients} & \textbf{Mean Years (SD)} && \textbf{Condition Pairs} & \textbf{N Patients} & \textbf{Mean Years (SD)} \\
\midrule
mental illness $\to$ reflux disorders & 385 & 6.7 (4.9) && hypertension $\to$ CKD & 435 & 7.2 (4.9) \\
epilepsy $\to$ mental illness & 287 & 7.6 (6.4) && mental illness $\leftrightarrow$ reflux disorders & 422 & 7.2 (5.9) \\
mental illness $\to$ insomnia & 267 & 5.9 (4.8) && hypertension $\to$ diabetes & 411 & 5.7 (4.5) \\
insomnia $\leftrightarrow$ reflux disorders & 216 & 4.3 (4.1) && diabetes $\to$ CKD & 381 & 5.6 (4.6) \\
cerebral palsy $\leftrightarrow$ epilepsy & 205 & 2.7 (4.1) && reflux disorders $\leftrightarrow$ chr. arthritis & 354 & 6.4 (5.3) \\
chr. airway diseases $\to$ insomnia & 167 & 6.9 (4.6) && chr. airway diseases $\to$ reflux disorders & 352 & 5.8 (5.1) \\
diabetes $\leftrightarrow$ hypertension & 149 & 3.8 (3.6) && hypertension $\to$ chr. arthritis & 340 & 6.9 (4.9) \\
mental illness $\leftrightarrow$ chr. pain conditions & 139 & 6.4 (4.8) && CKD $\leftrightarrow$ anaemia & 331 & 3.0 (3.4) \\
reflux disorders $\leftrightarrow$ IBD & 134 & 5.4 (4.5) && cardiac arrhythmias $\leftrightarrow$ CKD & 318 & 3.2 (3.9) \\
epilepsy $\to$ hypertension & 121 & 9.5 (6.1) && chr. airway diseases $\to$ chr. arthritis & 291 & 6.4 (5.0) \\
reflux disorders $\leftrightarrow$ chr. pain conditions & 121 & 5.0 (4.0) && reflux disorders $\to$ anaemia & 285 & 6.9 (5.4) \\
dysphagia $\leftrightarrow$ reflux disorders & 118 & 3.1 (3.5) && hypertension $\to$ cardiac arrhythmias & 279 & 7.5 (5.3) \\
reflux disorders $\to$ anaemia & 107 & 5.5 (4.5) && epilepsy $\to$ dysphagia & 256 & 9.5 (7.1) \\
epilepsy $\to$ chr. constipation & 95 & 6.1 (5.0) && hypertension $\to$ CHD & 256 & 6.2 (4.6) \\
insomnia $\to$ chr. arthritis & 92 & 6.1 (5.0) && cardiac arrhythmias $\to$ heart failure & 253 & 1.7 (3.0) \\
diabetes $\to$ CKD & 82 & 7.5 (6.8) && chr. airway diseases $\to$ cardiac arrhythmias & 245 & 7.2 (5.6) \\
hypertension $\leftrightarrow$ CKD & 82 & 7.4 (4.6) && CHD $\to$ CKD & 244 & 5.1 (4.5) \\
chr. pain conditions $\to$ neuropathic pain & 77 & 4.2 (5.2) && chr. airway diseases $\to$ CHD & 243 & 5.3 (5.3) \\
chr. airway diseases $\to$ cardiac arrhythmias & 76 & 7.8 (5.6) && chr. arthritis $\to$ cardiac arrhythmias & 242 & 7.1 (5.6) \\
cerebral palsy $\to$ dysphagia & 72 & 8.5 (6.5) && diabetes $\to$ cardiac arrhythmias & 239 & 5.9 (5.0) \\ \midrule
\multicolumn{3}{c}{\textbf{Females $<$ 45 years}} && \multicolumn{3}{c}{\textbf{Females $\geq$ 45 years}} \\
\cmidrule{1-3} \cmidrule{5-7}
\textbf{Condition Pairs} & \textbf{N Patients} & \textbf{Mean Years (SD)} && \textbf{Condition Pairs} & \textbf{N Patients} & \textbf{Mean Years (SD)} \\
\midrule
mental illness $\to$ reflux disorders & 375 & 7.4 (5.9) && hypertension $\to$ chr. arthritis & 484 & 6.3 (5.1) \\
chr. airway diseases $\to$ mental illness & 357 & 7.9 (6.0) && mental illness $\leftrightarrow$ reflux disorders & 473 & 6.2 (5.5) \\
chr. airway diseases $\to$ reflux disorders & 293 & 6.5 (5.3) && mental illness $\to$ chr. arthritis & 470 & 7.2 (5.7) \\
chr. pain conditions $\leftrightarrow$ mental illness & 260 & 5.7 (5.1) && hypertension $\to$ CKD & 462 & 7.0 (5.1) \\
chr. pain conditions $\to$ reflux disorders & 252 & 7.3 (5.2) && chr. arthritis $\leftrightarrow$ reflux disorders & 460 & 5.6 (5.1) \\
mental illness $\to$ insomnia & 240 & 6.2 (5.1) && mental illness $\leftrightarrow$ chr. airway diseases & 429 & 7.5 (6.2) \\
epilepsy $\to$ mental illness & 229 & 7.3 (5.9) && chr. arthritis $\to$ CKD & 405 & 6.9 (5.2) \\
chr. airway diseases $\to$ chr. pain conditions & 220 & 7.8 (5.9) && diabetes $\leftrightarrow$ hypertension & 405 & 4.8 (4.2) \\
reflux disorders $\leftrightarrow$ anaemia & 199 & 5.6 (5.0) && chr. airway diseases $\to$ chr. arthritis & 391 & 6.2 (5.2) \\
reflux disorders $\leftrightarrow$ insomnia & 179 & 5.5 (3.8) && chr. airway diseases $\to$ reflux disorders & 386 & 6.2 (5.1) \\
mental illness $\to$ diabetes & 167 & 6.8 (5.8) && menopausal \& perimenopausal $\leftrightarrow$ chr. arthritis & 365 & 5.8 (5.0) \\
chr. airway diseases $\to$ insomnia & 164 & 8.7 (5.8) && diabetes $\to$ CKD & 361 & 5.7 (4.5) \\
epilepsy $\to$ thyroid disorders & 160 & 7.4 (5.3) && mental illness $\to$ diabetes & 359 & 7.8 (6.0) \\
epilepsy $\leftrightarrow$ cerebral palsy & 158 & 8.0 (6.1) && reflux disorders $\to$ anaemia & 338 & 5.1 (4.6) \\
mental illness $\to$ chr. arthritis & 154 & 8.9 (6.3) && chr. arthritis $\to$ anaemia & 328 & 5.9 (5.2) \\
mental illness $\to$ IBD & 152 & 7.7 (4.7) && chr. airway diseases $\to$ diabetes & 300 & 5.5 (5.1) \\
mental illness $\to$ neuropathic pain & 147 & 7.2 (5.3) && thyroid disorders $\to$ Dementia & 286 & 8.5 (5.8) \\
chr. airway diseases $\to$ diabetes & 136 & 7.8 (5.6) && chr. arthritis $\to$ cardiac arrhythmias & 278 & 6.7 (5.5) \\
chr. airway diseases $\to$ chr. arthritis & 132 & 7.6 (5.9) && hypertension $\to$ cardiac arrhythmias & 267 & 8.2 (5.6) \\
epilepsy $\to$ dysphagia & 130 & 9.5 (6.3) && diabetes $\to$ anaemia & 267 & 6.1 (4.2) \\ \midrule
\multicolumn{6}{p{6in}}{
Arrows indicate temporal associations: $\to$ suggests the first condition typically precedes the second, while $\leftrightarrow$ indicates no preferred order. The total number of patients (N Patients) represents the number of individuals with each co-morbidity pair, regardless of order. Mean Years $\pm$ SD shows the average time between the occurrences of the two conditions, regardless of order. Chr. = chronic. CKD = chronic kidney disease. CHD = coronary heart disease.}\\
\bottomrule
\end{tabular}
\end{table}

\subsection*{Analysis of Trajectory Clusters}

\begin{figure}[]
    \centering
    \fbox{%
    \parbox{\textwidth}{%
    \centering
    \includegraphics[width=0.98\textwidth]{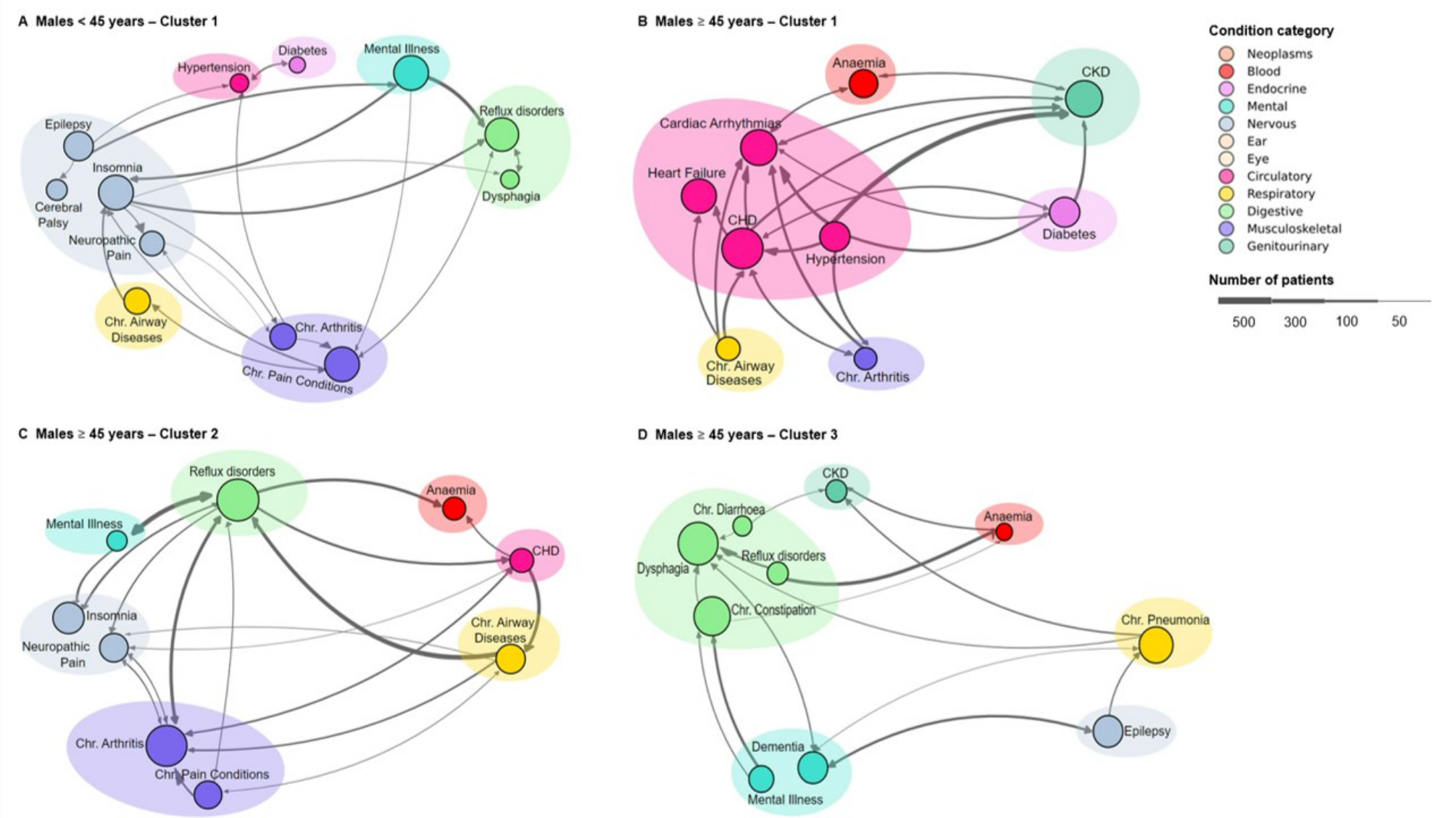}
    \vspace{0.5em}
    \hrule
    \vspace{0.5em} 
    \includegraphics[width=0.98\textwidth]{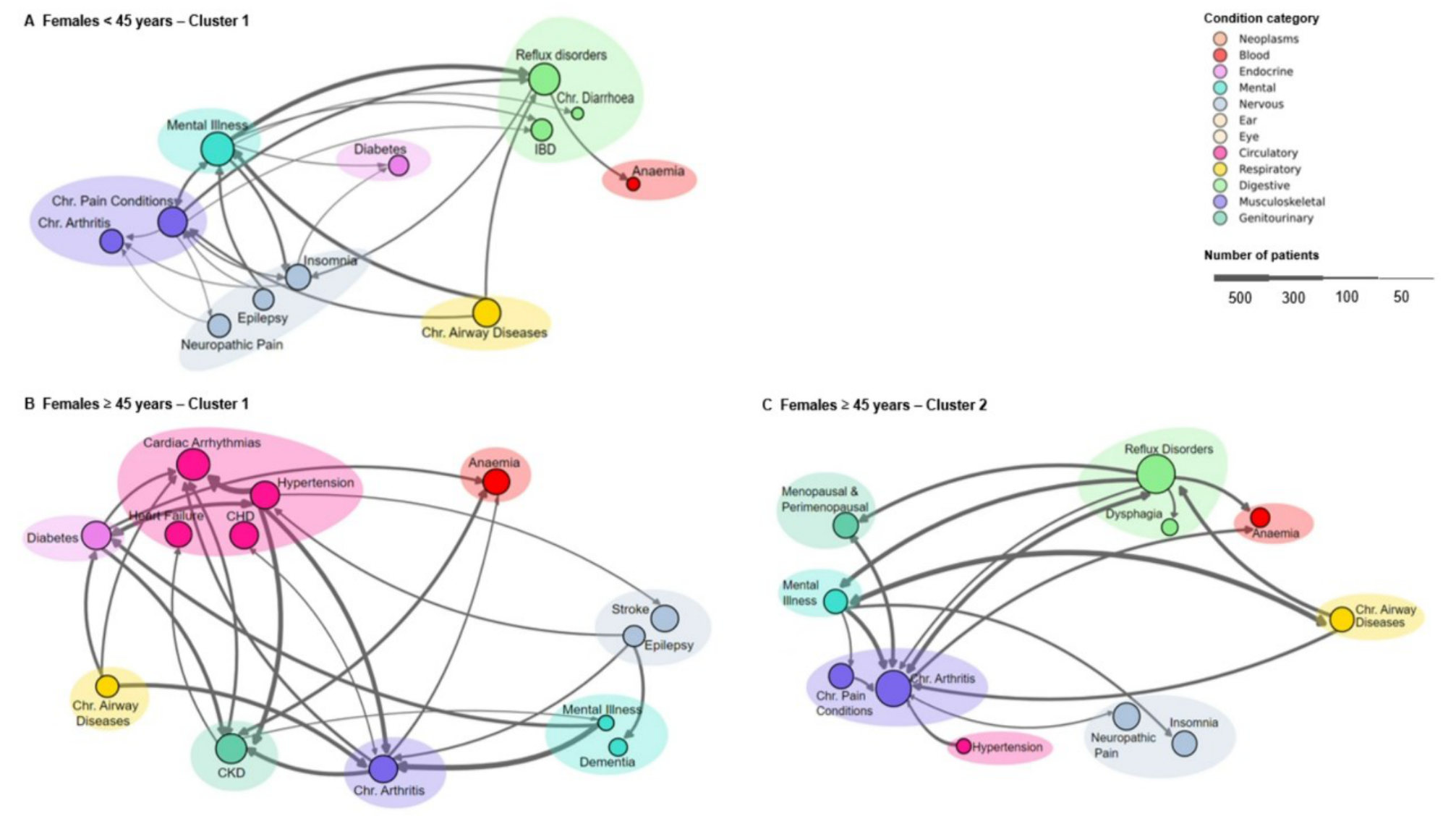}
    \caption{Disease trajectory clusters across age groups and genders. Top: Males - A) $<$ 45 years cluster 1, B) $\geq$ 45 years cluster 1, C) $\geq$ 45 years cluster 2, and D) $\geq$ 45 years cluster 3. Bottom: Females - A) $<$ 45 years single cluster;  B) $\geq$ 45 years cluster 1, and C) $\geq$ 45 years cluster 2. Node size indicates condition prevalence, with larger nodes representing more frequent occurrences. Conditions in each cluster represent more than 5\% of the total conditions. Edges show associations between conditions, with edge thickness corresponding to the frequency of condition pair occurrences (minimum edge frequency of 1\%). Coloured shaded areas group related conditions within the same category as defined in the legend. The legend specifies condition categories and provides a scale for patient numbers.} \label{fig:genderclusters}
    }
    }
\end{figure}

\paragraph*{Identified Clusters in Trajectories of Males} \noindent In the male sub-population, one cluster was identified for males under 45 years ($<$45) and three clusters for males 45 years and older ($\geq$45) (Table \ref{tab:cluster-scores}). Thirty-seven shared trajectories were considered for clustering in males $<$45 years, and 229 shared trajectories in males $\geq$45 years (Table~\ref{tab:ntraj}). Only trajectories with a minimum of ten patients were included. Table~\ref{tab:system-clusters-combined} presents an overview of the clusters identified within the male population.\\ 

\begin{table}[!h]
\centering
\tiny
\caption{Long-term condition clusters for A) males and B) females categorised by age groups. Each cluster presents the most frequent system condition categories included in the trajectories alongside the count of trajectories (N traj) and total patient numbers (N patients). System percentages (\%) are calculated based on the total number of trajectories in each cluster. Clusters are presented in descending order of patient count within each age category.}
\label{tab:system-clusters-combined}
\begin{tabular}{@{}p{1.5cm}p{1cm}p{1.5cm}p{6.5cm}p{1.5cm}p{1.5cm}@{}}
\toprule
\textbf{Cluster} & \textbf{N traj} & \textbf{N patients} & \textbf{System condition groups distribution (\%)} & \textbf{Mortality \%$^{a}$} & \textbf{Long hospital stay \%$^{a,b}$} \\
\midrule
\multicolumn{6}{l}{\textbf{A. Males}} \\
\midrule
\multicolumn{6}{l}{\textit{a. $<$ 45 years}} \\
\midrule
Cluster 1 & 37 & 549 & nervous (32.4\%), musculoskeletal (19.8\%), digestive (18.0\%), mental (12.6\%), respiratory (7.2\%), circulatory (3.7\%), endocrine (2.7\%), genitourinary (1.8\%), blood (1.8\%) & 16.9& 46.8\\
\midrule
\multicolumn{6}{l}{\textit{b. $\geq$ 45 years}} \\
\midrule
Cluster 1 & 112 & 2824 & circulatory (51.8\%), genitourinary (11.9\%), endocrine (8.0\%), blood (7.1\%), respiratory (7.1\%), musculoskeletal (4.8\%), nervous (4.8\%), digestive (4.5\%) &63.5 &84.1 \\
Cluster 2 & 81  & 1557 & musculoskeletal (25.1\%), digestive (20.6\%), nervous (18.1\%), circulatory (14.4\%), respiratory (8.6\%), blood (5.3\%), mental (4.5\%), ear (3.4\%) & 47.8&67.4 \\
Cluster 3 & 36  & 633  & digestive (39.8\%), mental (15.7\%), nervous (13.9\%), respiratory (13.9\%), circulatory (5.6\%), genitourinary (4.6\%), blood (2.8\%), ear (1.9\%), endocrine (1.9\%) &69.4 &82.2 \\
\midrule
\multicolumn{6}{l}{\textbf{B. Females}} \\
\midrule
\multicolumn{6}{l}{\textit{a.$<$ 45 years}} \\
\midrule
Cluster 1 & 88 & 1713 & digestive (24.6\%), nervous (21.6\%), musculoskeletal (18.6\%), mental (14.4\%), respiratory (10.2\%), endocrine (5.3\%), blood (2.3\%), circulatory (1.5\%), genitourinary (1.5\%) &13.4 &43.8 \\
\midrule
\multicolumn{6}{l}{\textit{b. $\geq$ 45 years}} \\
\midrule
Cluster 1 & 256 & 6101 & circulatory (34.1\%), nervous (12.1\%), musculoskeletal (10.5\%), genitourinary (8.9\%), endocrine (8.5\%), respiratory (6.9\%), blood (6.6\%), mental (5.5\%), digestive (3.9\%), neoplasms (2.2\%), ear (0.8\%) &58.5 &74.9 \\
Cluster 2 & 183  & 4057 & digestive (25.9\%), musculoskeletal (21.9\%), nervous (16.2\%), genitourinary (7.8\%), circulatory (7.7\%), respiratory (6.7\%), mental (6.6\%), blood (4.0\%), ear (1.6\%), endocrine (1.5\%), neoplasms (0.2\%) &43.0 &61.8 \\
\bottomrule
\multicolumn{6}{l}{\textsuperscript{a} Percentages (\%) are calculated using the unique number of patients included in each cluster.} \\
\end{tabular}
\end{table}

\noindent \textbf{For males under 45 years.} The analysis of shared trajectories in males revealed distinct patterns across different age groups (Figure \ref{fig:genderclusters}). For males under 45 years, neurological conditions were the most common affecting 32.4\% of this subpopulation. This was followed by conditions of the musculoskeletal (19.8\%) and digestive system (18.0\%). Mental health conditions were also notable, present in 12.6\% of cases. Looking at more specific conditions of this younger cohort revealed an equal prevalence (37.8\%) of chronic pain conditions, mental illness, and insomnia. Other significant conditions included reflux disorders (35.1\%), epilepsy (27.0\%), and chronic airway diseases (21.6\%) (Tables~\ref{tab:condition-male-clusters} and \ref{tab:mtrajless45}).\\\\
\textbf{For males 45 years and older.} The analysis of health trajectories in males aged 45 years and older revealed three distinct clusters, each characterised by unique patterns of multimorbidity. The largest cluster, encompassing 112 shared trajectories and 2824 patients, was predominantly defined by circulatory system conditions (51.8\%) (Table~\ref{tab:system-clusters-combined}). Coronary heart disease (CHD) emerged as the most prevalent condition, affecting 44.6\% of patients, closely followed by CKD (35.7\%) and cardiac arrhythmias (34.8\%) (Table~\ref{tab:condition-male-clusters} and \ref{tab:mtrajabove45}). Heart failure was also significantly present, occurring in 32.1\% of cases. This cluster highlighted a strong interplay between cardiovascular and renal health, further complicated by metabolic conditions such as diabetes and hypertension, each present in 24.1\% of patients. The presence of peripheral vascular disease (19.6\%) and chronic airway diseases (15.2\%) in this cluster underscores the complex associations between cardiovascular, respiratory, and metabolic systems in ageing men.\\
The second cluster, comprising 81 trajectories and 1557 patients, was dominated by musculoskeletal (25.1\%) and digestive system (20.6\%) conditions. Reflux disorders were most prevalent, affecting 53.1\% of patients, closely followed by chronic arthritis at 50.6\% (Table~\ref{tab:condition-male-clusters}). This cluster was prominently associated with the high incidence of chronic pain conditions (23.5\%) and neuropathic pain (24.7\%), indicating that chronic pain significantly affects many older men. Cardiovascular conditions remained relevant, with CHD affecting 17.3\% of patients. The presence of mental illness (12.3\%) and hearing loss (8.6\%) in this cluster points to the diverse health challenges faced by this subgroup.\\
The third cluster, though smaller with 36 trajectories and 633 patients was characterised by a high prevalence of digestive (39.8\%) and neurological (13.9\%) conditions. Dysphagia was the most common condition, present in 50.0\% of patients, followed closely by chronic constipation at 41.7\% (Table~\ref{tab:condition-male-clusters}). Neurological conditions were prominent, with both epilepsy and dementia affecting 27.8\% of patients each. The significant presence of chronic pneumonia (36.1\%) in this cluster highlights the persistent relevance of respiratory conditions in older males, while the occurrence of CKD (13.9\%) further emphasises the complex interplay of multiple organ systems in ageing.

\paragraph*{Identified Clusters in Trajectories of Females}
\noindent In the female sub-population, three clusters were obtained, one for females under 45 years ($<$45) and two for females 45 years and older ($\geq$45) (Table \ref{tab:cluster-scores}). Eighty-eight shared trajectories were considered for clustering in females $<$45 years, and 439 shared trajectories in females $\geq$45 years (Table~\ref{tab:ntraj}). Table~\ref{tab:system-clusters-combined} and Figure~\ref{fig:genderclusters} illustrates the distinct patterns of health trajectory clusters identified within the female population across different age groups.

\noindent \textbf{For females under 45 years.} The analysis of shared trajectories in females revealed distinct patterns across different age groups. For females under 45 years, digestive system conditions emerged as the predominant health concern, affecting 24.6\% of this subpopulation. This was closely followed by neurological conditions (21.6\%) and musculoskeletal conditions(18.6\%). Mental health conditions were also prominent, present in 14.4\% of cases. Looking at more specific conditions of this younger cohort revealed a high prevalence of mental illness (43.2\%), followed by reflux disorders (38.6\%), and chronic pain conditions (34.1\%). Other significant conditions included chronic airway diseases (30.7\%), insomnia (25.0\%), and chronic arthritis (21.6\%) (Table~\ref{tab:female-condition-clusters} and \ref{tab:ftrajbelow45}).

\noindent \textbf{For females 45 years and older.} The analysis of health trajectories in females aged 45 years and older revealed two distinct clusters.  The larger of these clusters, encompassing 256 shared trajectories and 6101 patients, was predominantly defined by circulatory system conditions (34.1\%) (Table~\ref{tab:system-clusters-combined}). Cardiac arrhythmias emerged as the most prevalent condition, affecting 30.5\% of patients, closely followed by CKD (26.6\%) and diabetes (24.2\%) (Table~\ref{tab:female-condition-clusters}). Hypertension was also significantly present, occurring in 23.4\% of cases. This cluster highlighted a strong interplay between cardiovascular and metabolic health, further complicated by musculoskeletal conditions such as chronic arthritis, present in 22.7\% of patients. The presence of CHD (22.3\%) and chronic airway diseases (14.5\%) in this group underscores the complex associations between cardiovascular, respiratory, and metabolic systems in ageing women. \\
A distinct health profile emerged in the other cluster, comprising 183 trajectories and 4057 patients, dominated by digestive (25.9\%) and musculoskeletal (21.9\%) conditions. Reflux disorders were most prevalent, affecting 48.1\% of patients, closely followed by chronic arthritis at 41.5\% (Table~\ref{tab:female-condition-clusters} and \ref{tab:ftrajabove45}). This group was significant for its high incidence of neuropathic pain (23.5\%) and chronic pain conditions (20.2\%), indicating that chronic pain significantly affects many older women. Menopausal and perimenopausal conditions were also prominent, affecting 20.8\% of patients. 

\begin{figure}[t]
    \centering
    \includegraphics[width=1.0\textwidth]{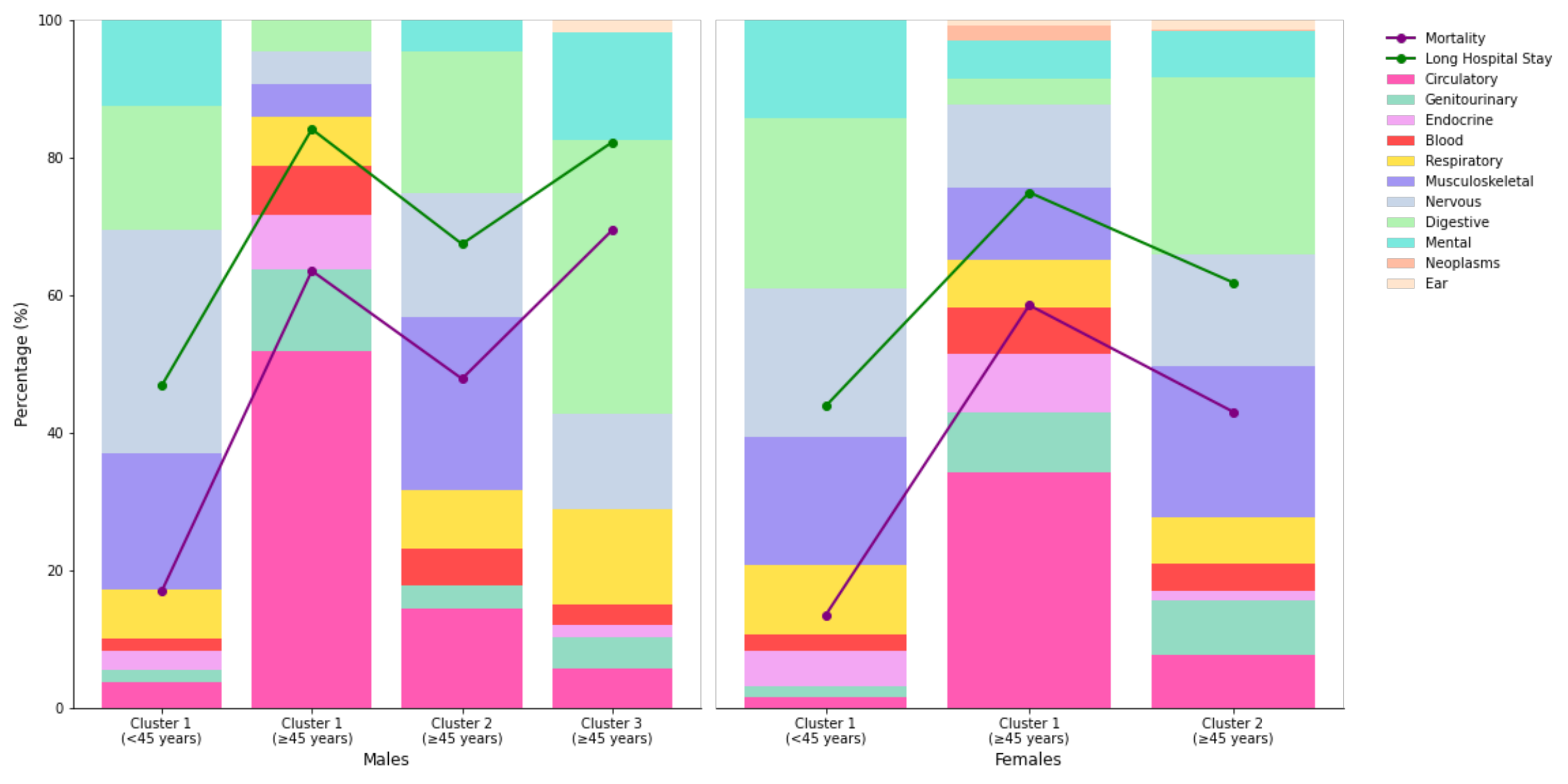}
    \caption{Distribution of long-term conditions and rates of mortality and long hospital stays across patient clusters.}
    \label{fig:mort-hosp}
\end{figure}

\paragraph*{Cluster Characteristics by Sex and Age Groups}
\noindent Figure ~\ref{fig:mort-hosp} and Table~\ref{tab:cluster_characteristics} present the mortality and long hospital stay ($\geq$ 4 days) rates observed across the identified clusters, stratified by sex and age groups. In the younger age group ($<$45 years), both males and females exhibited single clusters with similar mean ages (35.6 $\pm$ 5.8 and 35.9 $\pm$ 6.1 years, respectively). Mortality percentages for clusters $<$ 45 years were comparably low, with males at 16.9\% and females at 13.4\%. Long hospital stays were less common in this age group, with rates of 46.8\% for males and 43.8\% for females, significantly lower than in $\geq$ 45 year populations.\\
For the age group $\geq$45 years, we observed more distinct patterns in mortality and long hospital stays. In males, clusters 1 and 3 showed particularly high mortality percentages (63.5\% and 69.4\%). These clusters also had the highest rates of long hospital stays (84.1\% and 82.2\%, respectively). Cluster 1 was predominantly characterised by circulatory system conditions (51.8\%), while cluster 3 was notable for digestive system conditions (39.8\%), suggesting a strong association between these condition categories and both increased mortality risk and extended hospitalisations. Similarly, in females $\geq$45 years, we observed a similar pattern, with the larger cluster (cluster 1) showing a higher mortality rate (3.5 per 100 patient-years) and percentage (58.5\%) compared to cluster 2 (2.4 per 100 patient-years and 43.0\%). Cluster 1 also had a higher rate of long hospital stays (74.9\% vs 61.8\% in cluster 2) and was primarily affected by circulatory system conditions (34.1\%).

\begin{figure}[t]
     \includegraphics[width=1.0\textwidth]{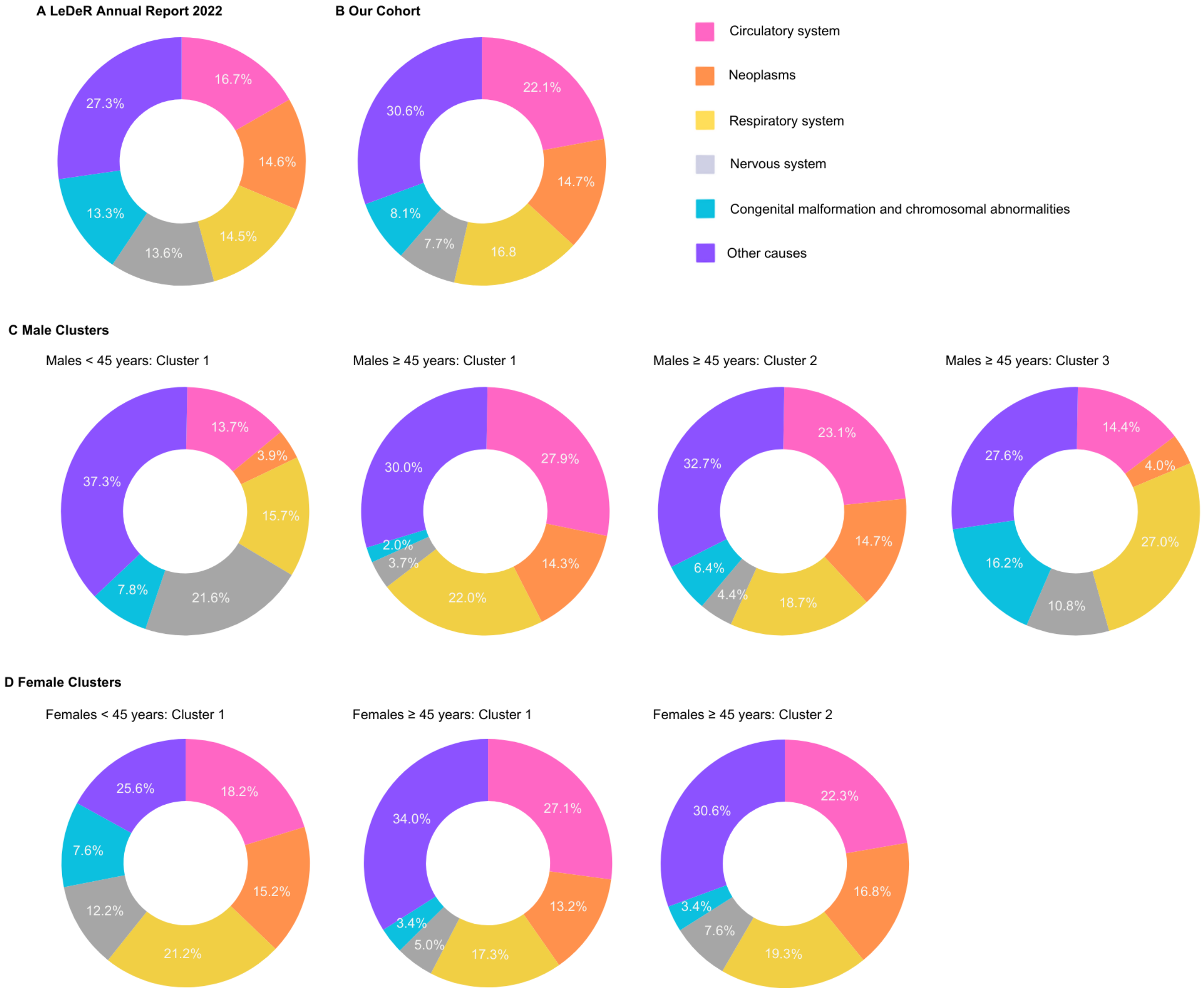}
         \caption{Comparison of the five leading causes of death among individuals with intellectual disabilities (ID), grouped based on ICD-10 chapters. Data presented includes findings from the A) LeDeR Annual Report 2022 \citep{white2023learning}, B) our cohort study using the SAIL (Secure Anonymised Information Linkage) databank and stratification by clusters based on sex and age groups for C) male cluster ($<$ 45 years: cluster 1 and $\geq$ 45 years: clusters 1, 2, 3) and D) females ($<$ 45 years: cluster 1; $\geq$ 45 years: clusters 1, 2). Values represent percentages for each cause of death.}
         \label{fig:causeofdeath}
\end{figure}

\noindent \textbf{Analysis of causes of death.} Figure~\ref{fig:causeofdeath} presents the distribution of the five leading causes of death among individuals with ID in our study population, which we can compare to the findings from the most recent annual report \textit{Learning from Lives and Deaths - people with a learning disability and autistic people} (LeDeR) \cite{white2023learning}. Across all groups, circulatory system conditions consistently appear as a major cause of death, particularly prominent in the older age groups ($\geq$45 years). Notably, in the $<$45 years age groups, neoplasms and respiratory system conditions feature more prominently as causes of death compared to the older groups. In the older age groups, we observe some variations between clusters, with cluster 1 for both males and females showing a higher proportion of deaths due to circulatory system conditions compared to other clusters. This aligns with our earlier observation of cluster 1 being characterised by a higher prevalence of circulatory system conditions. These findings underscore the complex interplay between age, sex, and specific health conditions in determining mortality patterns among individuals with ID.
\section{Discussion}
This study presents a novel approach for the identification and temporal analysis of LTC trajectories in adults with ID using both primary and secondary care data. By employing an unsupervised clustering method on trajectories of a large dataset of electronic health records, we have uncovered distinct patterns of co-morbidity and disease trajectories that vary significantly across sex and age groups. Our findings reveal a high prevalence of MLTCs in the ID population, with 85.9\% of individuals experiencing two or more long-term conditions. This aligns with previous research highlighting the complex health needs of this population \cite{cooper2020rates, mann2023scoping}. On average this group has 4.5 long-term conditions per patient.

\noindent \textbf{Sex differences:} Our analysis demonstrated significant sex differences in the prevalence of LTCs among adults with ID, while also identifying common patterns. Mental illness and epilepsy were the most prevalent conditions across both sexes, with mental illness affecting more than 30\% and epilepsy affecting approximately 30\% of both males and females. This prevalence aligns with previous findings in the literature \cite{carey2016health}. However, clear sex differences were apparent in other conditions. Females exhibited a higher prevalence of thyroid disorders (30.8\%) compared to males (14.1\%), and anaemia affected 20\% of females compared to 14\% of males. Females also showed increased prevalence of endocrine, skeletal, and digestive conditions, a finding consistent with previous research \cite{yang2022biomarkers}. Our investigation of temporal co-morbidities revealed that some conditions displayed opposite temporal associations in males and females for certain pairs of conditions. For instance, we observed that in females under 45 years, epilepsy tended to precede thyroid disorders, while this association was not significant in males of any age group.
This finding highlights the importance of investigating sex-stratified disease trajectories, as the temporal direction between conditions can vary depending on sex \cite{westergaard2019population}.

\noindent \textbf{Age-related patterns:} Age-related differences were apparent in condition patterns between $<$ 45 years and $\geq$ 45 years adults with ID. Under 45 groups showed a predominance of neurological (such as epilepsy and dementia), digestive (including dysphagia and chronic constipation), and mental health conditions, consistent with previous research in the ID population \cite{tyrer2019multimorbidity,van2017patterns}. In males under 45 years, neurological conditions accounted for 32.4\% of the identified cluster, while digestive system conditions were most prevalent (24.6\%) in females of the same age group. \\
In the over 45 groups, circulatory system conditions emerged as a primary concern for both sexes with ID. We found a higher prevalence of CHD, cardiac arrhythmias, heart failure, hypertension, kidney disease, and diabetes in older adults with ID. The co-occurrence of these conditions is particularly noteworthy, as it mirrors the complex interplay of cardiovascular and renal health observed in the general population \cite{de2022hypertension}. Furthermore, the strong association between hypertension and various other cardiovascular complications, including heart failure, arrhythmias, and CHD, aligns with patterns seen in the broader population \cite{masenga2023hypertensive}.

\noindent \textbf{Mortality and hospitalisation:} Across all age groups and clusters, males generally exhibited slightly higher mortality rates and percentages compared to females - a pattern observed in the general population but appears to be more pronounced in individuals with ID \cite{tyrer2022health}. Our analysis revealed a strong association between circulatory system conditions and both increased mortality and extended hospitalisations, particularly in older adults, highlighting these conditions as the most prevalent cause of death in the identified clusters. While mental health and neurological conditions are prevalent across all groups, their impact on mortality appears less direct compared to circulatory and certain digestive system conditions. However, mental conditions significantly contribute to extended hospital care needs and overall quality of life, especially in younger adults with ID \cite{siddiqui2018hospital}. We observed a clear association between higher rates of long hospital stays and increased mortality in specific clusters of the older population. For instance, in males $\geq$ 45 years, cluster 1 exhibited both the highest rate of long stays (84.1\%) and a high mortality rate  (3.9 per 100 patient-years). Lastly, our study underscores the value of temporality in predicting both co-morbidity and mortality \cite{beck2016diagnosis}.
 
\noindent \textbf{Limitations:} The reliance on clinical records may result in underdiagnosis of certain conditions, particularly those with subtle presentations in the ID population. Moreover, our analysis was constrained by limited ethnic diversity in the dataset. 

\noindent \textbf{Conclusion:} In conclusion, this study presents a data-driven overview of MLTC trajectory patterns in adults with ID. By revealing distinct clusters of conditions and their progression over time, our findings underscore the complex interplay of age, sex, and health conditions in adults with ID and provide a foundation for more targeted, personalised healthcare strategies. Our work advances understanding of how MLTCs manifest and progress in people with ID, revealing distinct patterns of disease development and complex interactions between conditions in this population. Future research aims to validate these clusters in diverse populations and investigate the underlying mechanisms of the observed progression patterns.

\section*{Acknowledgments}
The work was funded by National Institute for Health Research. The project is entitled “DECODE: Data-driven machinE-learning aided stratification and management of multiple long-term COnditions in adults with intellectual disabilitiEs.” Grant no.NIHR203981.

\section*{Conflict of Interest Statement} The authors declare that the research was conducted in the absence of any commercial or financial relationships that could be construed as a potential conflict of interest.

\section*{Author Contributions}
Writing original draft: R.K.; Conceptualisation and design: R.K, G.C.; Data preparation: R.K., E.A., A.A., F.Z., G.C., S.G.; Data curation: R.K., A.A., F.Z., S.G.; Funding acquisition: S.G., G.J., G.C., A.A., F.Z.; Analysis and modelling: R.K.; Interpretation of data: R.K., G.C.; Writing review and editing: R.K., G.C., E.A., R-z.K., F.Z., A.A., S.G., G.J.; Approving final version of manuscript: all authors.
\section*{Data Availability Statement}
The anonymised individual-level data sources used in this study are available in the SAIL Databank at Swansea University,
Swansea, UK, but as restrictions apply, they are not publicly available. All proposals to use SAIL data are subject to review by
the independent Information Governance Review Panel (IGRP). Before any data can be accessed, approval must be given by
the IGRP. The IGRP gives careful consideration to each project to ensure proper and appropriate use of SAIL data. When
access has been granted, it is gained through a privacy-protecting safe haven and remote access system referred to as the SAIL
Gateway. SAIL has established an application process to be followed by anyone who would like to access data via SAIL at:
\url{https://www.saildatabank.com/application-process/}.

\section*{Ethics Statement} The project was reviewed and approved by the SAIL Databank's independent Information Governance Review Panel (IGRP Project: 1375), ensuring compliance with ethical standards and data protection regulations. 


\newpage
\section*{Supplementary Material}

\renewcommand{\thefigure}{S\arabic{figure}}
\setcounter{figure}{0} 

\renewcommand{\thetable}{S\arabic{table}}
\setcounter{table}{0} 
\begin{center}
{
\small
\setlength{\tabcolsep}{2pt}
\begin{longtable}{llll}
\caption{Long-term conditions used in the study and their system categorisation.}
\label{tab:ltc_list}\\
\toprule
\textbf{Condition} & \textbf{Abbreviated Name} &  \textbf{Included Conditions} &\textbf{Condition Category} \\
\midrule
\endfirsthead
\multicolumn{4}{c}{\tablename\ \thetable\ -- \textit{Continued from previous page}} \\
\toprule
Condition & Abbreviated Name & Included Conditions& Category \\
\endhead
\hline
\multicolumn{4}{r}{\textit{Continued on next page}} \\
\endfoot
\hline
\endlastfoot
      Addisons Disease & & &Endocrine \\
      Anaemia & & &Blood \\
      Barretts Oesophagus& & & Digestive \\
      Bronchiectasis & & &Respiratory \\
      Cancer & & &Neoplasms \\
      Cardiac Arrhythmias & & \parbox[t]{4cm}{Atrial Fibrillation, Atrial Flutter, Supraventricular Tachycardia, Heart Block}& Circulatory \\
      Cerebral Palsy & & &Nervous \\
      Chronic Airway Diseases & Chr. Airway Diseases &\parbox[t]{3.5cm}{Eosinophil Bronchitis, Chronic Obstructive Pulmonary Disease (COPD), Asthma} & Respiratory \\
\end{longtable}
}
\end{center}

\newpage
\begin{center}
{
\setlength{\tabcolsep}{2pt}
\begin{longtable}{>{\raggedright}p{3.5cm} rrr}
\caption{Number of individuals with 1 to $\geq$ 6 conditions by sex.}
\label{tab:nconditions-gender} \\
\toprule
\textbf{Number of conditions} & \textbf{All Patients} & \textbf{Males} & \textbf{Females} \\
\midrule
\endfirsthead
\multicolumn{4}{c}%
{\tablename\ \thetable\ -- \textit{Continued from previous page}} \\
\toprule
\textbf{Condition Category} & \textbf{Total} & \textbf{MALE} & \textbf{FEMALE} \\
\midrule
\endhead
\midrule
\multicolumn{4}{r}{\textit{Continued on next page}} \\
\endfoot
\bottomrule
\endlastfoot
1 condition & 1838 & 1073 & 765 \\
2 conditions & 2104 & 1261 & 843 \\
3 conditions & 1955 & 1056 & 899 \\
4 conditions & 1784 & 936 & 848 \\
5 conditions & 1393 & 718 & 675 \\
6+ conditions & 3995 & 1786 & 2209 \\
\end{longtable}
}
\end{center}
\newpage
\begin{center}
{
\setlength{\tabcolsep}{2pt}
\begin{longtable}{p{3.5cm}p{1.5cm}p{1cm}} 
\caption{Distribution of males and females across age, ethnicities, and deprivation quintiles groups.}
\label{tab:gndr-demo} \\
\toprule
\textbf{Category} & \textbf{Males} & \textbf{Females} \\
\midrule
\endfirsthead
\multicolumn{3}{c}%
{\tablename\ \thetable\ -- \textit{Continued from previous page}} \\
\toprule
\textbf{Category} & \textbf{Male} & \textbf{Female} \\
\midrule
\endhead
\midrule
\multicolumn{3}{r}{\textit{Continued on next page}} \\
\endfoot
\bottomrule
\endlastfoot
\textbf{Total} & 6830 & 6239 \\
\midrule
\multicolumn{3}{l}{\textbf{Age, years}} \\
$<$45 & 3010 & 2496 \\
$\geq$ 45 & 3820 & 3743 \\
\midrule
\multicolumn{3}{l}{\textbf{Ethnic group}} \\
Asian & 104 & 75 \\
Black & 26 & 17 \\
Mixed & 9 & 6 \\
Other & 16 & 17 \\
Unknown & 1904 & 1734 \\
White & 4771 & 4390 \\
\midrule
\multicolumn{3}{l}{\textbf{WIMD 2019 Quintile}}\\

1. Most deprived & 1659 & 1536\\
2 & 1345 & 1237 \\
3 & 1084 & 1003 \\
4 & 1022 & 927\\
5. Least deprived & 679 & 667 \\
\end{longtable}
}
\end{center}

\newpage
\begin{longtable}{@{}l*{6}{r@{\qquad}r}@{}}
\caption{Condition counts for all males and by age group.} 
\label{tab:malecondcounts} \\
\toprule
\multirow{2}{*}{Condition} & \multicolumn{2}{c}{All Males} & \multicolumn{2}{c}{Males $<$ 45} & \multicolumn{2}{c}{Males $\geq$ 45} \\
\cmidrule(lr){2-3} \cmidrule(lr){4-5} \cmidrule(lr){6-7}
& Count & \% & Count & \% & Count & \% \\
\midrule
\endfirsthead

\multicolumn{7}{c}{Table \thetable{} continued} \\
\toprule
\multirow{2}{*}{Condition} & \multicolumn{2}{c}{All Males} & \multicolumn{2}{c}{Males $<$ 45} & \multicolumn{2}{c}{Males $\geq$ 45} \\
\cmidrule(lr){2-3} \cmidrule(lr){4-5} \cmidrule(lr){6-7}
& Count & \% & Count & \% & Count & \% \\
\midrule
\endhead

\midrule
\multicolumn{7}{r}{Continued on next page} \\
\endfoot

\bottomrule
\endlastfoot

Mental Illness & 2239 & 32.8 \qquad & 1054 & 35.0 \qquad & 1185 & 31.0 \\
Epilepsy & 2120 & 31.0 \qquad & 1062 & 35.3 \qquad & 1058 & 27.7 \\
Reflux Disorders & 2027 & 29.7 \qquad & 894 & 29.7 \qquad & 1133 & 29.7 \\
Hypertension & 1602 & 23.5 \qquad & 467 & 15.5 \qquad & 1135 & 29.7 \\
Chr. Airway Diseases & 1597 & 23.4 \qquad & 690 & 22.9 \qquad & 907 & 23.7 \\
Diabetes & 1383 & 20.2 \qquad & 415 & 13.8 \qquad & 968 & 25.3 \\
CKD & 1342 & 19.6 \qquad & 313 & 10.4 \qquad & 1029 & 26.9 \\
Chr. Arthritis & 1264 & 18.5 \qquad & 331 & 11.0 \qquad & 933 & 24.4 \\
\end{longtable}

\newpage
\begin{longtable}{@{}l*{6}{r@{\qquad}r}@{}}
\caption{Condition Counts for all females and by age group.} 
\label{tab:femalecondcounts} \\
\toprule
\multirow{2}{*}{Condition} & \multicolumn{2}{c}{All Females} & \multicolumn{2}{c}{Females $<$ 45} & \multicolumn{2}{c}{Females $\geq$ 45} \\
\cmidrule(lr){2-3} \cmidrule(lr){4-5} \cmidrule(lr){6-7}
& Count & \% & Count & \% & Count & \% \\
\midrule
\endfirsthead
\multicolumn{7}{c}{Table \thetable{} continued} \\
\toprule
\multirow{2}{*}{Condition} & \multicolumn{2}{c}{All Females} & \multicolumn{2}{c}{Females $<$ 45} & \multicolumn{2}{c}{Females $\geq$ 45} \\
\cmidrule(lr){2-3} \cmidrule(lr){4-5} \cmidrule(lr){6-7}
& Count & \% & Count & \% & Count & \% \\
\midrule
\endhead
\midrule
\multicolumn{7}{r}{Continued on next page} \\
\endfoot
\bottomrule
\endlastfoot

Mental Illness & 2188 & 35.1 \qquad & 932 & 37.3 \qquad & 1256 & 33.6 \\
Thyroid Disorders & 1921 & 30.8 \qquad & 676 & 27.1 \qquad & 1245 & 33.3 \\
Reflux Disorders & 1876 & 30.1 \qquad & 772 & 30.9 \qquad & 1104 & 29.5 \\
Epilepsy & 1766 & 28.3 \qquad & 781 & 31.3 \qquad & 985 & 26.3 \\
Chr. Airway Diseases & 1698 & 27.2 \qquad & 714 & 28.6 \qquad & 984 & 26.3 \\
Hypertension & 1496 & 23.9 \qquad & 298 & 11.9 \qquad & 1194 & 31.9 \\
Chr. Arthritis & 1471 & 23.6 \qquad & 300 & 12.0 \qquad & 1171 & 31.3 \\
CKD & 1313 & 21.0 \qquad & 286 & 11.5 \qquad & 1027 & 27.4 \\
\end{longtable}

\newpage

\begin{longtable}{l@{\qquad}*{10}{r@{\qquad}}}
\caption{Comparison of Calinski-Harabasz scores for different numbers of clusters in spectral clustering \label{tab:cluster-scores}} \\
\hline
& \multicolumn{10}{c}{Number of clusters} \\
\cline{2-11}
& 2 & 3 & 4 & 5 & 6 & 7 & 8 & 9 & 10 \\
\hline
\endfirsthead
\multicolumn{11}{c}{(Continued from previous page)} \\
\hline
& \multicolumn{10}{c}{Number of clusters} \\
\cline{2-11}
& 2 & 3 & 4 & 5 & 6 & 7 & 8 & 9 & 10 \\
\hline
\endhead
\hline
\multicolumn{11}{r}{(Continued on next page)} \\
\endfoot
\hline
\endlastfoot
Males \\
\quad $<$45 years & 13.7 \qquad & 9.6 \qquad & 6.1 \qquad & 6.2 \qquad & 4.8 \qquad & 4.4 \qquad & 3.2 \qquad & 3.8 \qquad & 3.3 \\
\quad $\geq$ 45 years & 82.1 \qquad & 103.8 \qquad & 66.2 \qquad & 62.2 \qquad & 46.1 \qquad & 45.8 \qquad & 42.9 \qquad & 41.9 \qquad & 38.2 \\
Females \\
\quad $<$45 years & 15.1 \qquad & 14.8 \qquad & 6.5 \qquad & 6.2 \qquad & 5.8 \qquad & 4.9 \qquad & 4.2 \qquad & 4.8 \qquad & 4.2 \\
\quad $\geq$  45 years & 116.3 \qquad & 115.9 \qquad & 98.9 \qquad & 77.4 \qquad & 70.4 \qquad & 62.2 \qquad & 57.0 \qquad & 47.8 \qquad & 46.5 \\
\end{longtable}

\begin{longtable}{@{}lcc@{}}
\caption{Number of trajectories (N traj) considered in the clustering algorithm for a trajectory length of three, stratified by sex and age group. Only trajectories with a minimum of ten patients were included.\label{tab:ntraj}}\\
\toprule
N traj & Males & Females \\
\midrule
\endfirsthead
\multicolumn{3}{l}{Table \thetable{} continued from previous page} \\
\toprule
N traj & Males & Females \\
\midrule
\endhead
\midrule
\multicolumn{3}{r}{Continued on next page} \\
\endfoot
\bottomrule
\endlastfoot
$<$45 years & 37 & 88 \\
$\geq$45 years & 229 & 439 \\
\end{longtable}
\newpage

\newpage

\begin{longtable}{@{}p{1.5cm}p{2cm}p{10cm}@{}}
\caption{Condition percentages of the trajectories included in clusters for males, stratified by age groups. Each cluster presents the
most frequent conditions including in the trajectories alongside the count of trajectories (N traj) and total patient numbers (N patients). Percentages (\%) are calculated based on the total number of trajectories in
each cluster. Clusters are presented in descending order of patient count within each age category. \label{tab:condition-male-clusters}}\\
\toprule
\textbf{N traj} & \textbf{N patients} & \textbf{Condition percentages and clusters} \\
\midrule
\endfirsthead
\multicolumn{3}{l}{Table \thetable{} continued from previous page} \\
\toprule
N traj & N patients & Condition percentages and clusters \\
\midrule
\endhead
\midrule
\multicolumn{3}{r}{Continued on next page} \\
\endfoot
\bottomrule
\endlastfoot
\multicolumn{3}{l}{\textbf{a. Males $<$ 45 years}} \\
\midrule
37& 549& chr. pain conditions (37.8\%), mental illness (37.8\%), insomnia (37.8\%), reflux disorders (35.1\%), epilepsy (27.0\%), chr. airway diseases (21.6\%), chr. arthritis (21.6\%), neuropathic pain (18.9\%), cerebral palsy (13.5\%), dysphagia (10.8\%), hypertension (10.8\%), diabetes (8.1\%), CKD (5.4\%), chr. constipation (5.4\%), anaemia (5.4\%) \\
\midrule
\multicolumn{3}{l}{\textbf{b. Males $\geq$ 45 years}} \\
\midrule
112& 2824& CHD (44.6\%), CKD (35.7\%), cardiac arrhythmias (34.8\%), heart failure (32.1\%), diabetes (24.1\%), hypertension (24.1\%), anaemia (21.4\%), PVD (19.6\%), chr. airway diseases (15.2\%), chr. arthritis (13.4\%), stroke (7.1\%), reflux disorders (6.3\%), chr. pneumonia (6.3\%), Cancer (5.4\%) \\
\midrule
81& 1557& reflux disorders (53.1\%), chr. arthritis (50.6\%), insomnia (29.6\%), chr. airway diseases (25.9\%), neuropathic pain (24.7\%), chr. pain conditions (23.5\%), CHD (17.3\%), anaemia (16.0\%), mental illness (12.3\%), cardiac arrhythmias (9.9\%), heart failure (9.9\%), hearing loss (8.6\%) \\
\midrule
36& 633& dysphagia (50.0\%), chr. constipation (41.7\%), chr. pneumonia (36.1\%), epilepsy (27.8\%), dementia (27.8\%), mental illness (19.4\%), reflux disorders (13.9\%), CKD (13.9\%), chr. diarrhoea (11.1\%), anaemia (8.3\%), chr. airway diseases (5.6\%), cardiac arrhythmias (5.6\%), cerebral palsy (5.6\%), thyroid disorders (5.6\%), stroke (5.6\%) \\
\end{longtable}

\newpage
\begin{longtable}{@{}llp{12cm}@{}}
\caption{Condition percentages of the trajectories included in clusters for females, stratified by age. Each cluster presents the
most frequent conditions including in the trajectories alongside the count of trajectories (N traj) and total patient numbers (N patients). Percentages (\%) are calculated based on the total number of trajectories in
each cluster. Clusters are presented in descending order of patient count within each age category. \label{tab:female-condition-clusters}}\\
\toprule
\textbf{N traj} & \textbf{N patients} & \textbf{Condition percentages and clusters} \\
\midrule
\endfirsthead
\multicolumn{3}{l}{Table \thetable{} continued from previous page} \\
\toprule
\textbf{N traj} & \textbf{N patients} & \textbf{Condition percentages and clusters} \\
\midrule
\endhead
\midrule
\multicolumn{3}{r}{Continued on next page} \\
\endfoot
\bottomrule
\endlastfoot
\multicolumn{3}{l}{\textbf{a. Females $<$ 45 years}} \\
\midrule
88& 1713&  mental illness (43.2\%), reflux disorders (38.6\%), chr. pain conditions (34.1\%), chr. airway diseases (30.7\%), insomnia (25.0\%), chr. arthritis (21.6\%), neuropathic pain (20.5\%), IBD (18.2\%), epilepsy (17.0\%), diabetes (15.9\%), dysphagia (8.0\%), anaemia (6.8\%), chr. diarrhoea (5.7\%) \\\\
\midrule
\multicolumn{3}{l}{\textbf{b. Females $\geq$ 45 years}} \\
\midrule
256& 6101& cardiac arrhythmias (30.5\%), CKD (26.6\%), diabetes (24.2\%), hypertension (23.4\%), chr. arthritis (22.7\%), CHD (22.7\%), stroke (21.1\%), heart failure (19.9\%), anaemia (19.9\%), chr. airway diseases (14.5\%), epilepsy (13.3\%), dementia (9.4\%), mental illness (9.4\%), cancer (6.6\%), chr. pneumonia (6.3\%), PVD (5.9\%), chr. pain conditions (5.1\%), dysphagia (5.1\%) \\
\midrule
183& 4057& reflux disorders (48.1\%), chr. arthritis (41.5\%), neuropathic pain (23.5\%), menopausal \& perimenopausal (20.8\%), chr. pain conditions (20.2\%), insomnia (20.2\%), chr. airway diseases (19.1\%), mental illness (18.6\%), anaemia (12.0\%), dysphagia (8.7\%), IBD (7.7\%), hypertension (7.1\%), chr. constipation (6.6\%), chr. diarrhoea (6.6\%) \\ \\
\end{longtable}

\newpage
\begin{center}
    {
    \small
\setlength{\tabcolsep}{2pt}
\begin{longtable}{|l|l|l|c|c|c|c|}
\caption{The shared LTC trajectories of length three identified within the male sub-population below 45 years of age. The column N Patients indicates the total number of patients that share this trajectory following this chronological order among the conditions.}
\label{tab:mtrajless45}\\
\hline
\textbf{Condition 1} & \textbf{Condition 2} & \textbf{Condition 3} & \textbf{Cluster} & \textbf{N Patients} & \textbf{Mortality \%} & \textbf{Long Hospital Stay \%} \\ \hline
\endfirsthead

\multicolumn{7}{c}%
{{\bfseries \tablename\ \thetable{} -- continued from previous page}} \\
\hline
\textbf{Condition 1} & \textbf{Condition 2} & \textbf{Condition 3} & \textbf{Cluster} & \textbf{N Patients} & \textbf{Mortality \%} & \textbf{Long Stay \%} \\ \hline
\endhead

\hline \multicolumn{7}{r}{{Continued on next page}} \\ \hline
\endfoot

\hline
\endlastfoot

mental illness & insomnia & reflux disorders & 1 & 41 & 14.6 & 29.3 \\ \hline
epilepsy & mental illness & insomnia & 1 & 30 & 18 & 50 \\ \hline
mental illness & insomnia & chronic arthritis & 1 & 23 & 20 & 34.8 \\ \hline
chronic airway diseases & insomnia & reflux disorders & 1 & 23 & 0 & 26.1 \\ \hline
epilepsy & mental illness & reflux disorders & 1 & 21 & 0 & 47.6 \\ \hline
epilepsy & cerebral palsy & dysphagia & 1 & 18 & 27 & 72 \\ \hline
chronic airway diseases & insomnia & chronic arthritis & 1 & 18 & 0 & 38.9 \\ \hline
mental illness & chronic pain conditions & reflux disorders & 1 & 18 & 0 & 44.4 \\
\hline

\end{longtable}
}
\end{center}

\newpage

\begin{center}
    {
    \small
\setlength{\tabcolsep}{2pt}
\begin{longtable}{|l|l|l|c|c|c|c|}
\caption{The shared LTC trajectories of length three identified within the male sub-population aged 45 years and older. The column N Patients indicates the total number of patients that share this trajectory following this chronological order among the conditions.}
\label{tab:mtrajabove45}\\
\hline
\textbf{Condition 1} & \textbf{Condition 2} & \textbf{Condition 3} & \textbf{Cluster} & \textbf{N Patients} & \textbf{Mortality \%} & \textbf{Long Hospital Stay \%} \\ \hline
\endfirsthead

\multicolumn{7}{c}%
{{\bfseries \tablename\ \thetable{} -- continued from previous page}} \\
\hline
\textbf{Condition 1} & \textbf{Condition 2} & \textbf{Condition 3} & \textbf{Cluster} & \textbf{N Patients} & \textbf{Mortality \%} & \textbf{Long Hospital Stay \%} \\ \hline
\endhead

\hline \multicolumn{7}{r}{{Continued on next page}} \\ \hline
\endfoot

\hline
\endlastfoot

hypertension & diabetes & ckd & 1 & 77 & 61 & 77.9 \\ \hline
hypertension & ckd & cardiac arrhythmias & 1 & 55 & 70.9 & 87.3 \\ \hline
hypertension & cardiac arrhythmias & heart failure & 1 & 54 & 63 & 87 \\ \hline
hypertension & ckd & heart failure & 1 & 53 & 75.5 & 92.5 \\ \hline
diabetes & ckd & anaemia & 1 & 53 & 62.3 & 86.8 \\ \hline
hypertension & ckd & anaemia & 1 & 50 & 72 & 88 \\ \hline
chronic airway diseases & cardiac arrhythmias & heart failure & 1 & 49 & 63.3 & 91.8 \\ \hline
diabetes & ckd & cardiac arrhythmias & 1 & 47 & 66 & 91.5 \\
\hline

\end{longtable}
}
\end{center}

\newpage
\begin{center}
    {
    \small
\setlength{\tabcolsep}{2pt}
\begin{longtable}{|l|l|l|c|c|c|c|}
\caption{The shared LTC trajectories of length three identified within the female sub-population aged less than 45 years. The column N Patients indicates the total number of patients that share this trajectory following this chronological order among the conditions.}
\label{tab:ftrajbelow45}\\
\hline
\textbf{Condition 1} & \textbf{Condition 2} & \textbf{Condition 3} & \textbf{Cluster} & \textbf{N Patients} & \textbf{Mortality \%} & \textbf{Long Hospital Stay \%} \\ \hline
\endfirsthead

\multicolumn{7}{c}%
{{\bfseries \tablename\ \thetable{} -- continued from previous page}} \\
\hline
\textbf{Condition 1} & \textbf{Condition 2} & \textbf{Condition 3} & \textbf{Cluster} & \textbf{N Patients} & \textbf{Mortality \%} & \textbf{Long Hospital Stay \%} \\ \hline
\endhead

\hline \multicolumn{7}{r}{{Continued on next page}} \\ \hline
\endfoot

\hline
\endlastfoot

chronic airway diseases & mental illness & reflux disorders & 1 & 55 & 0 & 27.3 \\ \hline
mental illness & chronic pain conditions & reflux disorders & 1 & 47 & 10.6 & 38.3 \\ \hline
chronic airway diseases & chronic pain conditions & reflux disorders & 1 & 46 & 13 & 39.1 \\ \hline
chronic airway diseases & mental illness & chronic pain conditions & 1 & 37 & 0 & 35.1 \\ \hline
chronic airway diseases & mental illness & insomnia & 1 & 36 & 14.1 & 50 \\ \hline
epilepsy & mental illness & reflux disorders & 1 & 34 & 0 & 38.2 \\ \hline
mental illness & insomnia & chronic arthritis & 1 & 31 & 0 & 38.7 \\ \hline
chronic airway diseases & chronic pain conditions & chronic arthritis & 1 & 30 & 0 & 46.7 \\
\hline

\end{longtable}
}
\end{center}

\newpage
\begin{center}
    {
    \footnotesize
\setlength{\tabcolsep}{1pt}
\begin{longtable}{|l|l|l|c|c|c|c|}
\caption{The shared LTC trajectories of length three identified within the female sub-population aged 45 years and older. The column N Patients indicates the total number of patients that share this trajectory following this chronological order among the conditions}
\label{tab:ftrajabove45}\\
\hline
\textbf{Condition 1} & \textbf{Condition 2} & \textbf{Condition 3} & \textbf{Cluster} & \textbf{N Patients} & \textbf{Mortality \%} & \textbf{Long Hospital Stay \%} \\ \hline
\endfirsthead

\multicolumn{7}{c}%
{{\bfseries \tablename\ \thetable{} -- continued from previous page}} \\
\hline
\textbf{Condition 1} & \textbf{Condition 2} & \textbf{Condition 3} & \textbf{Cluster} & \textbf{N Patients} & \textbf{Mortality \%} & \textbf{Long Hospital Stay \%} \\ \hline
\endhead

\hline \multicolumn{7}{r}{{Continued on next page}} \\ \hline
\endfoot

\hline
\endlastfoot

hypertension & chronic arthritis & ckd & 1 & 77 & 54.5 & 74.0 \\ \hline
chronic airway diseases & chronic arthritis & ckd & 1 & 63 & 50.8 & 65.1 \\ \hline
diabetes & hypertension & ckd & 1 & 62 & 67.7 & 77.4 \\ \hline
hypertension & ckd & anaemia & 1 & 62 & 66.1 & 74.2 \\ \hline
hypertension & ckd & cardiac arrhythmias & 1 & 61 & 68.9 & 82.0 \\ \hline
hypertension & chronic arthritis & cardiac arrhythmias & 1 & 59 & 57.6 & 81.4 \\ \hline
hypertension & ckd & heart failure & 1 & 59 & 84.7 & 86.4 \\ \hline
chronic airway diseases & diabetes & ckd & 1 & 57 & 59.6 & 66.7 \\
\hline
\end{longtable}
}
\end{center}

\newpage

\begin{table}
\small
\setlength{\tabcolsep}{4pt}
\caption{Characteristics of clusters defined by shared disease trajectories. This table presents the distribution of unique patients shared all trajectories across identified clusters, stratified by sex and age groups ($<$ 45 and $\geq$ 45 years).}
\label{tab:cluster_characteristics}
\begin{tabular}{lc|ccc|c|cc}
\toprule
\multirow{2}{*}{} & \multicolumn{4}{c|}{Males} & \multicolumn{3}{c}{Females} \\
\cmidrule(lr){2-5} \cmidrule(lr){6-8}
& $<$ 45 years & \multicolumn{3}{c|}{$\geq$ 45 years} & $<$ 45 years & \multicolumn{2}{c}{$\geq$ 45 years} \\
\midrule
Clusters & 1 & 1 & 2 & 3 & 1 & 1 & 2 \\
\midrule
Total Patients & 549.0 & 2824.0 & 1557.0 & 633.0 & 1713.0 & 6101.0 & 4057.0 \\
Unique Patients & 301.0 & 728.0 & 525.0 & 320.0 & 491.0 & 1201.0 & 1009.0 \\
Mean age ($\pm$ SD) & 35.6 (5.8) & 62.7 (10.2) & 59.4 (9.6) & 59.7 (9.7) & 35.9 (6.1) & 63.0 (11.3) & 59.3 (10.8) \\
Autism, N (\%) & 26.0 (8.6) & 45.0 (6.2) & 26.0 (5.0) & 29.0 (9.1) & 30.0 (6.1) & 24.0 (2.0) & 22.0 (2.2) \\
Psychotropic Medications, N (\%) & 68.0 (22.6) & 81.0 (11.1) & 60.0 (11.4) & 69.0 (21.6) & 74.0 (15.1) & 171.0 (14.2) & 130.0 (12.9) \\
\bottomrule
\end{tabular}
\end{table}

\newpage

\begin{center}
\begin{figure}[!ht]
    \centering
    \includegraphics[]{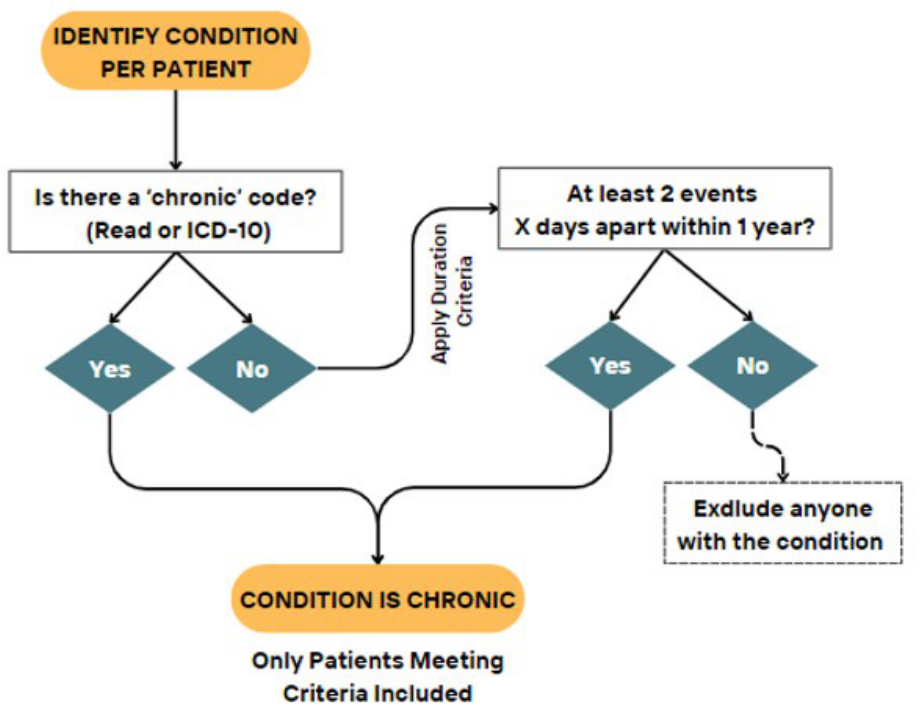}
    \caption{Defining chronic condition criteria.}
    \label{fig:duration}
\end{figure}
\end{center}
\newpage

\begin{figure}[!ht]
    \centering
    \includegraphics[width=0.75\textwidth]{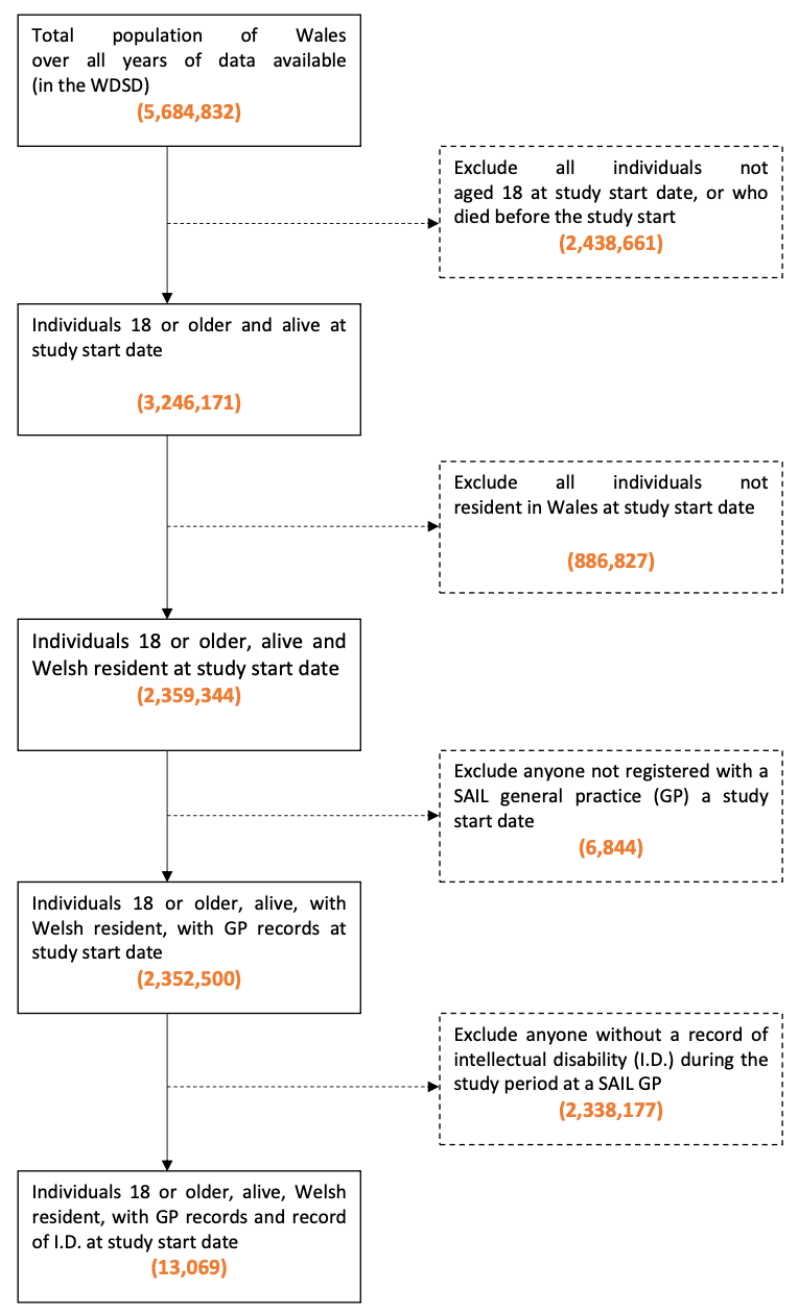}
    \caption{Consort Flow Diagram.}
    \label{fig:flowchart}
\end{figure}

\begin{figure}[!ht]
    \centering
    \includegraphics[width=\textwidth]{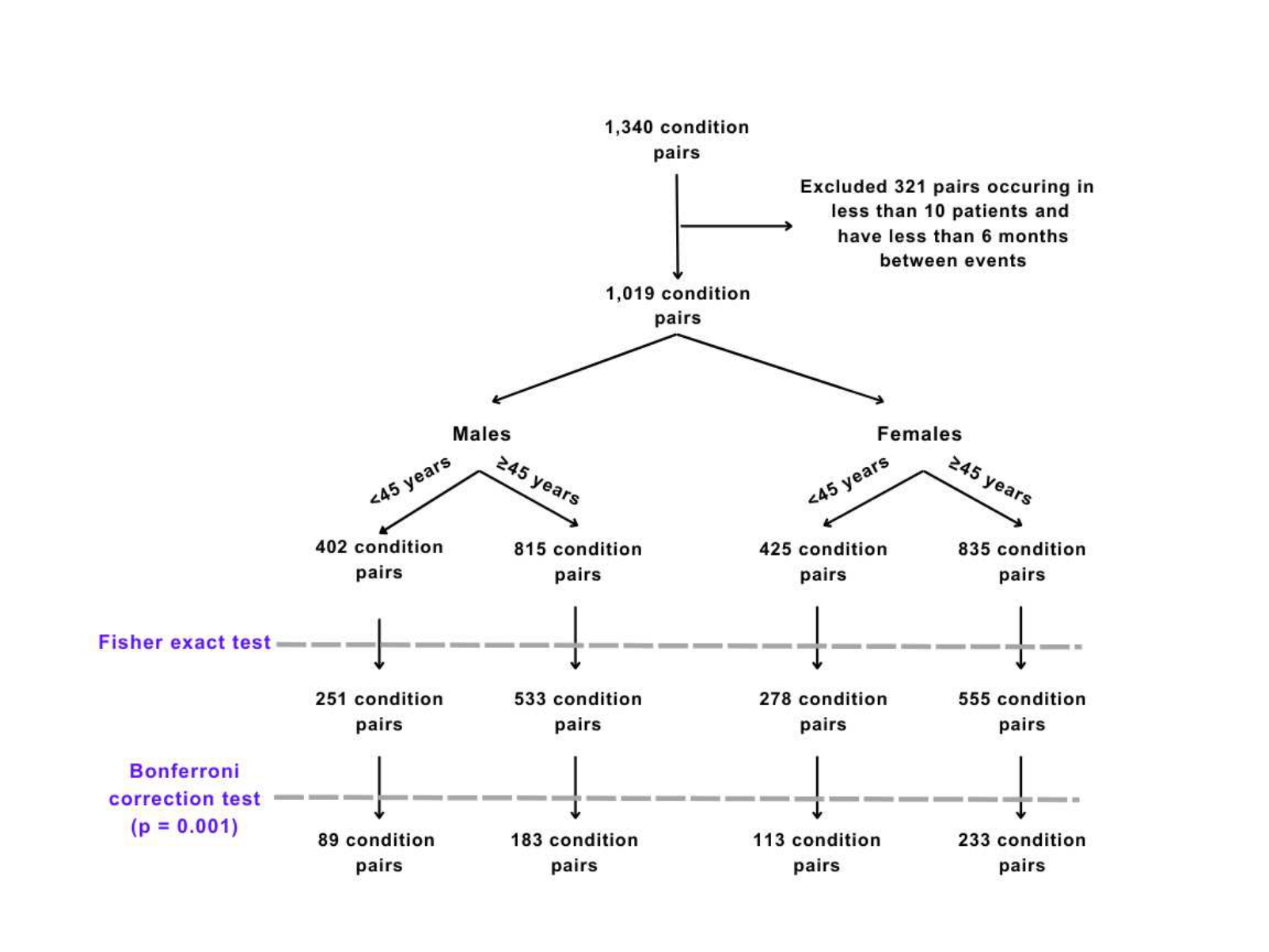}
    \caption{Long-term condition co-occurrences found in intellectual disability population data from 13,069 patients. 1,340 condition pairs were found to occur in the population; of these, a number of pairs were filtered out due to low frequency (N$<$10), or their events occur in less than six months period. After applying the Fisher exact test and Bonferroni correction the final condition pairs were found for both males and females stratified by age.}
    \label{fig:pairs-flowchart}
\end{figure}

\newpage
\bibliographystyle{unsrtnat}

\end{document}